\documentclass[sigconf]{acmart}
\AtBeginDocument{%
  }

\setcopyright{acmlicensed}
\copyrightyear{2025}
\acmYear{2025}
\acmDOI{XXXXXXX.XXXXXXX}
\acmConference[Conference acronym 'XX]{Make sure to enter the correct
  conference title from your rights confirmation email}{June 03--05,
  2018}{Woodstock, NY}
\acmISBN{978-1-4503-XXXX-X/2018/06}

\usepackage{amssymb}
\usepackage[most]{tcolorbox}
\usepackage{multirow}
\usepackage{balance}
\usepackage{tabularx}
\usepackage{booktabs} 
\usepackage{pifont}   

\begin{document}

\title{RAIR: A Rule-Aware Multimodal Benchmark for Challenging E-Commerce Relevance Assessment}

\author{Chenji Lu}
\authornote{Equally contribution.
\textsuperscript{\textdagger}Corresponding author.}
\email{luchenji.lcj@alibaba-inc.com}
\affiliation{%
  \institution{Taobao \& Tmall Group of Alibaba}
  \city{Beijing}
  \country{China}
}

\author{Zhuo Chen}
\authornotemark[1]
\email{cz462596@alibaba-inc.com}
\affiliation{%
  \institution{Taobao \& Tmall Group of Alibaba}
  \city{Beijing}
  \country{China}
}

\author{Hui Zhao}
\email{shuqian.zh@taobao.com}
\affiliation{%
  \institution{Taobao \& Tmall Group of Alibaba}
  \city{Beijing}
  \country{China}
}

\author{Zhenyi Wang}
\email{suge.wzy@taobao.com}
\affiliation{%
  \institution{Taobao \& Tmall Group of Alibaba}
  \city{Beijing}
  \country{China}
}

\author{Pengjie Wang}
\email{wangpengjie0421@163.com}
\affiliation{%
  \institution{Taobao \& Tmall Group of Alibaba}
  \city{Beijing}
  \country{China}
}

\author{Chuan Yu}
\email{yuchuan.yc@taobao.com}
\affiliation{%
  \institution{Taobao \& Tmall Group of Alibaba}
  \city{Beijing}
  \country{China}
}


\author{Jian Xu}
\authornotemark[2]
\email{xiyu.xj@taobao.com}
\affiliation{%
  \institution{Taobao \& Tmall Group of Alibaba}
  \city{Beijing}
  \country{China}
}


\renewcommand{\shortauthors}{Lu et al.}

\begin{abstract}
While Large Language Models (LLMs) have shown promise in e-commerce search, the field lacks benchmarks that offer both high complexity and standardized protocols. We introduce \textbf{RAIR} (\textbf{R}ule-\textbf{A}ware benchmark with \textbf{I}mage for \textbf{R}elevance), a large-scale Chinese benchmark derived from real-world industrial scenarios. Unlike prior datasets, RAIR employs a deterministic rule system to transform subjective judgments into objective protocols. We construct three strategic subsets: (1) a General Subset with industry-balanced sampling to evaluate fundamental capabilities; (2) a Hard Subset targeting reasoning-heavy and knowledge-dependent cases to probe performance upper bounds; and (3) a Visually Salient Subset for cases where visual evidence is strictly necessary for accurate judgment, enabling targeted evaluation of multimodal integration capabilities. Extensive experiments with 14 state-of-the-art models reveal that RAIR poses significant challenges: even GPT-5 achieves only 84.5\% binary accuracy and 43.3\% Macro-F1 on the General Subset. We release RAIR as both a standard industry reference and a testbed for assessing reasoning and instruction-following in LLMs and Visual Language Models (VLMs). Our data is available at \url{https://anonymous.4open.science/r/RAIR-2946/}.
\end{abstract}


\vspace{-0.2cm}

\ccsdesc[500]{Information systems~Relevance assessment}
\ccsdesc[500]{benchmark}
\ccsdesc[500]{Information systems~Large models}

\keywords{Web E-Commerce, Benchmark, Large Language Model, Search Relevance}
\begin{teaserfigure}
\centering
  \includegraphics[width=0.9\textwidth]{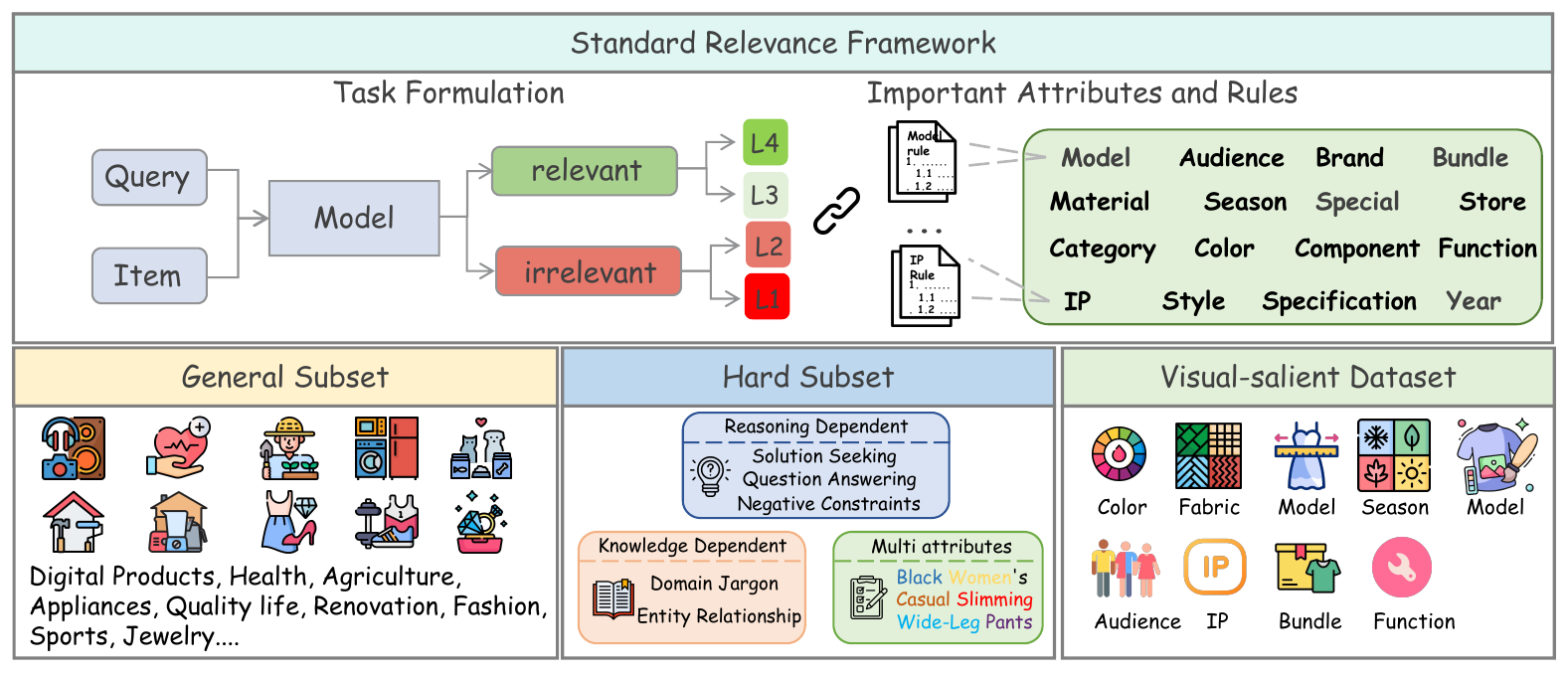}
  \vspace{-0.4cm}
  \caption{Overview of RAIR benchmark}
    \vspace{-0.1cm}
  \label{fig:head}
\end{teaserfigure}

\received{20 February 2007}
\received[revised]{12 March 2009}
\received[accepted]{5 June 2009}

\maketitle

\section{Introduction}
\label{sec:intro}
Search relevance stands as the cornerstone of modern e-commerce systems. On massive platforms like Taobao and Amazon, the core challenge lies in accurately aligning user queries with relevant items from billion-scale catalogs \cite{buy_2023,review_need1,review_need2}. As the fundamental determinant of user satisfaction and gross merchandise value (GMV), relevance serves as the critical prerequisite for all subsequent retrieval and ranking stages \cite{buy3, whybuy}.

\begin{figure}
    \label{fig:intro}
    \centering
    \includegraphics[width=1\linewidth]{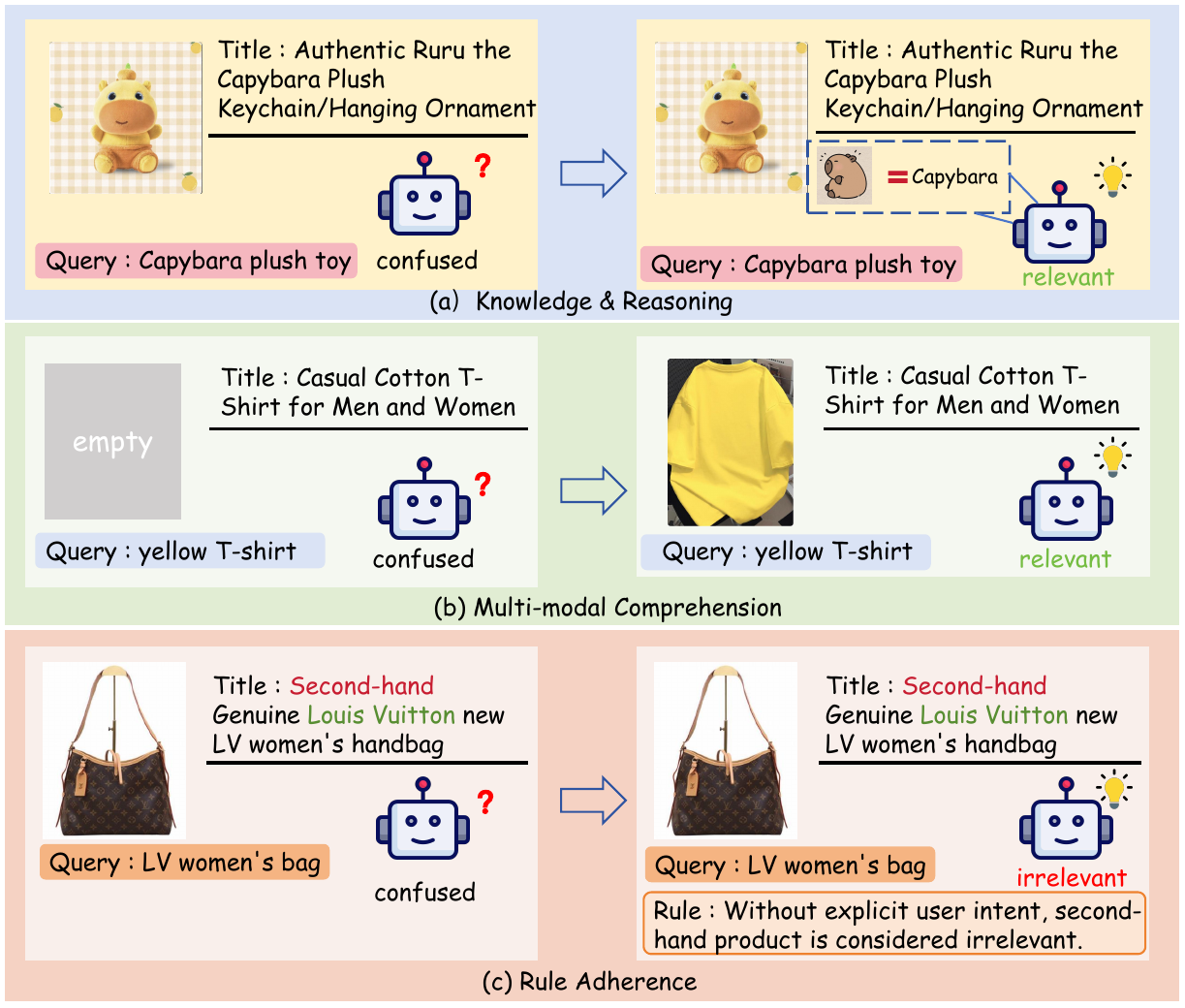}
    \caption{Overview of three core model capabilities evaluated in the RAIR benchmark.}
    \vspace{-0.6cm}
    \label{fig:intro}
\end{figure}
Driven by its pivotal role, search relevance has attracted extensive research attention \cite{huang2013learning,palangi2014semantic}. Recently, Large Language Models (LLMs) have emerged as the dominant backbone, demonstrating remarkable capabilities \cite{walmartllm,JD2,TaoSR1}. State-of-the-art methods place particular emphasis on leveraging Chain-of-Thought (CoT) reasoning to enhance accuracy. For instance, ELLM \cite{ELLM} decomposes relevance into a two-stage paradigm, performing attribute extraction prior to decision-making, while LREF \cite{LREF} further integrates synthesized multi-dimensional reasoning paths with rule references.

However, despite these modeling advancements, the community lacks a specialized benchmark to effectively evaluate such reasoning-driven capabilities. Existing datasets suffer from two critical limitations: 
\textbf{Insufficient Challenge:} Designed for broad multi-task understanding, benchmarks like Shopping MMLU \cite{shopingmmmu} lack dedicated mechanisms for mining relevance-specific hard cases, leading to performance saturation—e.g., even Qwen3-4B achieves 84.3\% accuracy on its relevance sub-task, while the 235B model scores only 81.8\%, failing to discriminate model capabilities across scales (detailed in Section~\ref{sec:6.3}).
\textbf{Lack of Standardized Reasoning Protocols:} Datasets like Shopping Queries \cite{shoppingqueries} provide coarse labels (e.g., Exact, Substitute) but lack a formalized rule system defining category boundaries. This absence of explicit adjudication logic creates a ``black box'' evaluation environment, severely hindering CoT-based models that rely on transparent reasoning steps. Consequently, there is an urgent need for a challenging benchmark grounded in a standardized reasoning framework.

To construct a challenging benchmark, we deconstruct the relevance task into three critical dimensions essential for modern relevance models, as illustrated in Figure \ref{fig:intro}.

(1) \textbf{Knowledge \& Reasoning Capability}~\cite{deepseek,skywork}.
Complex e-commerce scenarios demand reasoning that transcends basic keyword matching. As shown in Figure \ref{fig:intro}(a), accurate judgment often requires bridging the semantic gap between abstract queries and specific items using external world knowledge (e.g., mapping nicknames to scientific names) and logical inference.

(2) \textbf{Multi-modal Comprehension Capability}~\cite{multi-modal1,multi-modal2}.
In real-world search, visual cues are often indispensable for resolving textual ambiguities. As depicted in Figure \ref{fig:intro}(b), visual modalities provide critical discriminative features—such as specific styles or patterns—that text alone cannot capture. While current methods predominantly rely on LLMs, integrating visual understanding represents an inevitable frontier for robust relevance modeling.

(3) \textbf{Rule Adherence Capability.} 
Relevance judgments, inherently subjective, must be anchored in objective protocols. This dimension mandates strict adherence to sophisticated, expert-crafted rule systems (e.g., compatibility protocols shown in Figure~\ref{fig:intro}(c)). Fundamentally, this constitutes a specialized form of ``instruction following''~\cite{instruct1}, essential for ensuring reliable AI behavior.

Guided by these insights, we introduce \textbf{RAIR} (\textbf{R}ule-\textbf{A}ware benchmark with \textbf{I}mage for \textbf{R}elevance), a comprehensive benchmark designed to rigorously evaluate model competencies in e-commerce search. As depicted in Figure \ref{fig:head}, RAIR encompasses two synergistic components:

\textbf{1. A Standardized, Human-Aligned Framework.}
To resolve the ambiguity inherent in subjective labeling, we establish a deterministic adjudication protocol that codifies human preferences into explicit rules.
We define a fine-grained four-level scale (L1--L4) and deconstruct potential user query intents into 16 distinct attribute dimensions (e.g., Category, Brand).
Crucially, we construct a rigorous expert-validated system that deterministically maps the degree of attribute satisfaction to these specific relevance levels.
This comprehensive framework serves as a critical carrier, not only enabling models to deeply comprehend the nuances of relevance tasks but also providing a robust standard for evaluating their true relevance assessment capabilities.


\textbf{2. A Comprehensive Evaluation Benchmark.}
Grounded in this framework, we curated a dataset comprising three strategic subsets to probe specific model capabilities: 
(1) The \textbf{General Subset} targets fundamental competencies by enforcing extensive diversity. Instead of mimicking skewed real-world distributions, it utilizes stratified sampling to capture a wide spectrum of real-world scenarios across various industry sectors, ensuring a robust evaluation of model generalization.
(2) The \textbf{Hard Subset} stress-tests the model's upper boundaries in logic and knowledge. We target three specific high-complexity intent types—Reasoning-Dependent, Knowledge-Dependent, and Multi-Attribute constraints—and construct an automated mining pipeline that synergizes rule-based retrieval with LLM-based filtering to isolate the most challenging samples. 
(3) The \textbf{Visually Salient Subset} specifically assesses multimodal integration. We identify visual-dependency potential across 16 core attributes and leveraged LLMs to expand a lexicon of visual-specific terms. Candidates were then retrieved and rigorously filtered through a hybrid Rule+LLM pipeline to retain only those cases where visual evidence is strictly necessary for accurate judgment. 
Each instance in RAIR includes granular assessment criteria, establishing a new standard for e-commerce relevance evaluation.

Overall, our main contributions can be summarized as follows:
\begin{itemize}
    \item \textbf{Framework Standardization:} We establish a rigorous and comprehensive theoretical framework for e-commerce relevance assessment. By deconstructing user intents and defining deterministic mapping protocols, this system effectively transforms subjective relevance judgments into objective, reproducible standards.
    
    \item \textbf{Advanced Data Construction:} We design specialized automated pipelines that synergize rule-based retrieval with LLM-based filtering. This approach efficiently mines high-complexity samples—spanning reasoning-heavy, knowledge-dependent, and visually salient cases—from massive logs to construct challenging benchmark subsets.

    \item \textbf{The RAIR Benchmark:} We introduce RAIR, a large-scale multimodal dataset grounded in granular annotation rules. 
    Extensive experiments demonstrate RAIR's discriminative power and challenge.
    Beyond industrial application, RAIR serves as a rigorous testbed for broader AI capabilities—including complex reasoning and instruction following—driving future advancements in relevance modeling.
\end{itemize}

\section{Related Works}
\subsection{E-commerce Relevance Method}
Prior to the LLM era, relevance modeling was dominated by encoder-based architectures, primarily categorized into representation-based (dual-tower)~\cite{huang2013learning,palangi2014semantic,reimers2019sentence,twotower,DeepBoW,poly} and interaction-based (single-tower)~\cite{devlin2018bert,liu2019roberta,wang2019structbert} paradigms.
Recently, Large Language Models (LLMs) have transformed this field by leveraging extensive inherent knowledge~\cite{walmartllm}. A prevailing trend in modern approaches—such as ELLM~\cite{ELLM}, LREF~\cite{LREF}, and TaoSR1~\cite{TaoSR1}—is the utilization of CoT reasoning to decompose relevance decisions.
By explicitly incorporating discriminative logic into the inference process~\cite{DPO,deepseek}, these methods demonstrate that relevance modeling has evolved beyond simple classification. 
This paradigm shift highlights the foundational importance of a \textit{standardized and comprehensive rule system} to guide and evaluate the complex reasoning capabilities required for high-precision relevance assessment.

\vspace{-0.2cm}
\subsection{E-commerce Relevance Benchmark}
General e-commerce benchmarks have laid the foundation for domain comprehension. ECinstruct~\cite{Ecinstruct} and EcomGPT~\cite{eomgpt} focus on instruction tuning and atomic tasks, while eCeLLM~\cite{EceLLM} covers diverse tasks like sentiment analysis and sequential recommendation. Similarly, Amazon-M2~\cite{amazon} assesses capabilities such as concept understanding and multilingual processing. However, these works primarily serve as general comprehension benchmarks rather than specialized frameworks for granular relevance assessment.

Among benchmarks that incorporate relevance tasks, Shopping Queries~\cite{shoppingqueries} and Shopping MMLU~\cite{shopingmmmu} utilize the standard ESCI label space: \textit{Exact (E), Substitute (S), Complement (C),} and \textit{Irrelevant (I)}. EcomMMMU~\cite{EcomMMMU} adopts a similar four-tier taxonomy defined as: \textit{Fully Relevant (satisfying specifications), Functional Substitute, Combinable Complement,} and \textit{Irrelevant}. 
Critically, while these datasets define the label categories, they typically lack comprehensive annotation protocols or detailed judgment rules, leaving the specific criteria for each level ambiguous.

Regarding multimodal challenges, EcomMMMU's \textit{Visually Salient Subset} demonstrated the value of visual context and inspired our approach. However, a distinct methodological gap exists: their subset was constructed by filtering within existing annotated data, whereas our work tackles the challenge of actively \textit{mining} such complex cases directly from massive unlabeled logs.
This underscores the need for a benchmark that combines rigorous rule-based grounding with the ability to identify hard, visually dependent samples from the wild.

\vspace{-0.2cm}
\section{Standard Relevance Framework}
\subsection{Relevance Task}
\label{sec:3.1}
\begin{figure}[htbp]
    \centering
    \includegraphics[width=0.9\linewidth]{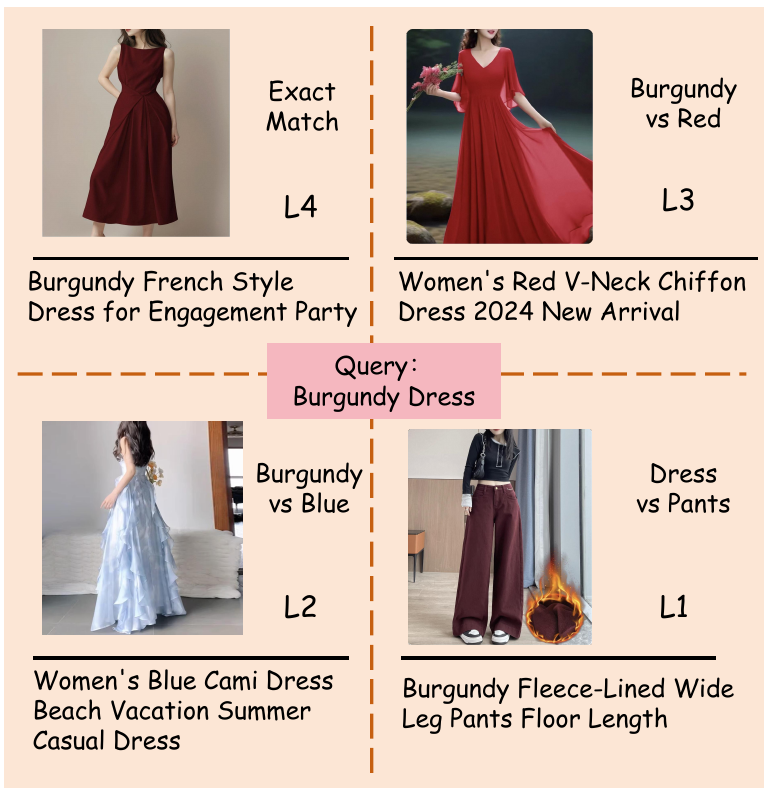}
    \vspace{-0.3cm}
    \caption{The fine-grained relevance hierarchy in RAIR. Using query ``Burgundy Dress,'' we illustrate the transition from category mismatch (L1) to perfect alignment (L4).}
    \vspace{-0.3cm}
    \label{fig:level}
\end{figure}

While e-commerce relevance is traditionally modeled as a binary task, we reformulate it as a fine-grained classification problem $f: (Q, I) \rightarrow y \in \{L1, L2, L3, L4\}$. As illustrated in Figure~\ref{fig:level} with the query \textit{``Burgundy Dress''}, the labels define a strict hierarchy of satisfaction: \textbf{L4 (Perfect Match)} represents ideal alignment; \textbf{L3 (Partial Match)} captures semantic proximity without explicit conflict (e.g., \textit{Red} vs. Burgundy); \textbf{L2 (Explicit Mismatch)} denotes specific attribute conflicts (e.g., \textit{Blue} vs. Burgundy); and \textbf{L1 (Completely Inconsistent)} indicates fundamental category errors (e.g., \textit{Pants} vs. Dress). This granularity is critical for optimization: \textbf{L3} provides a necessary boundary for ambiguous items that are neither clearly irrelevant nor fully satisfying, while separating \textbf{L1} from \textbf{L2} distinguishes severe user experience failures (category errors) from minor attribute mismatches.

\vspace{-0.3cm}
\subsection{Rule-Aware Relevance Assessment}
\label{sec:3.2}
In the domain of e-commerce, relevance evaluation relies on explicit protocols as a critical supplement to general world knowledge, rather than purely on semantic similarity. While proprietary standards exist within the industry, RAIR open-sources a comprehensive rule framework to standardize this evaluation. The goal is to align models strictly with human preferences, providing a transparent carrier to assess whether models can match—or even exceed—human-level judgment.

Aligning with this standard, we propose a Query-Centric framework that deconstructs user intent into 16 distinct attribute dimensions (e.g., Category, Brand), formally denoted as the Attributes Set $\mathcal{D}$. These dimensions were rigorously selected based on high-frequency analysis of industrial search logs and finalized through consensus among domain experts.
We constructed a rigorous system where expert-defined logic deterministically maps the degree of attribute satisfaction into a fine-grained L1--L4 spectrum (Figure~\ref{fig:head}). 
Crucially, this framework addresses boundary conditions through a severity-based priority mechanism. 
For instance, in a case where an item violates multiple dimensions—such as a ``red high heel'' for the query ``green shirt'' (mismatched in both \textit{Category} and \textit{Color})—the system resolves the label based on the most severe violation (e.g., assigning L1-Irrelevant due to the fundamental Category mismatch).
Validated by iterative expert reviews to ensure generalizability, detailed definitions of these attributes and protocols are in Appendix~\ref{app:B}.

Consequently, we redefine relevance judgment from an open-ended estimation to a determination strictly conditioned on explicit guidelines as follows:

\vspace{-0.3cm}
\begin{equation}
    y = f(Q, I \mid R)
\end{equation}
\vspace{-0.1cm}
where $R$ represents the applicable rule set that grounds the mapping from query intent to item attributes. The comprehensive documentation of this rule system will be open-sourced alongside the datasets.

\vspace{-0.1cm}
\section{Benchmark Construction} 

\begin{figure*}
    \centering
\includegraphics[width=1\linewidth]{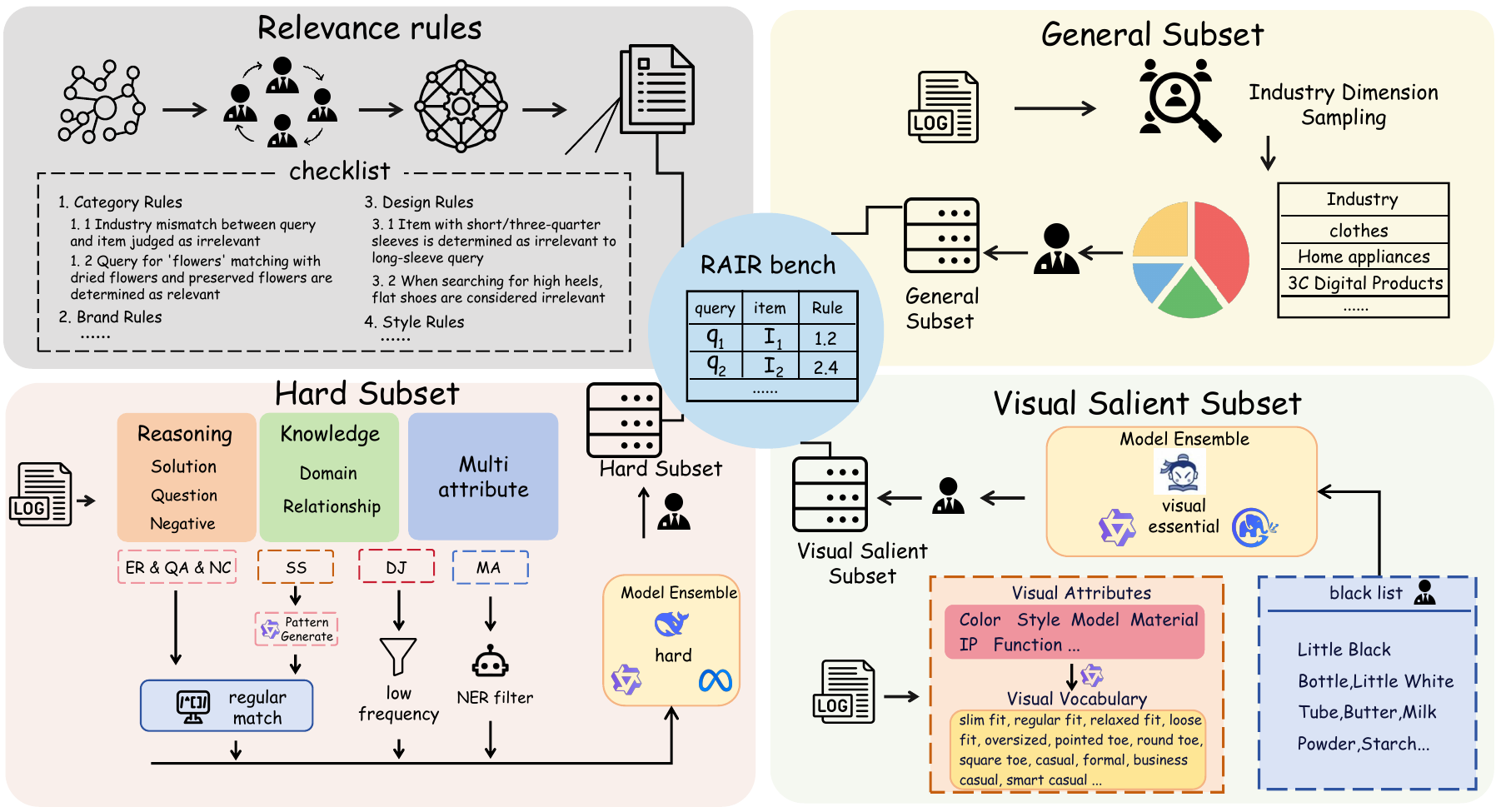}
    \caption{Construction process of RAIR benchmark, including the rule system composition and construction pipelines for General Subset, Hard Subset, and Visually Salient Subset.}
    \vspace{-0.3cm}
    \label{fig:main}
\end{figure*}

Grounded in the standardized framework established in Section~\ref{sec:3.2}—which serves as both the labeling protocol and the analytical lens for our data construction—we now detail the RAIR benchmark. 
Figure~\ref{fig:main} illustrates our pipeline, which originates from real-world Taobao user search logs and is engineered to capture diverse challenges through three distinct subsets: General, Hard, and Visually Salient (Sections~\ref{sec:general}--\ref{sec:visual}). 
Crucially, every sample is explicitly aligned with our rule system (Section~\ref{sec:rule-aware}). 
To ensure benchmark reliability, the entire corpus underwent rigorous professional annotation and multi-stage quality assurance, as detailed in Appendix~\ref{app:quality_control}.


\vspace{-0.3cm}
\subsection{General Subset}
\label{sec:general}
The General Subset is designed to evaluate foundational e-commerce capabilities—including reasoning, domain knowledge, and attribute extraction—across a broad spectrum of product categories. To achieve this, maintaining a diverse and balanced distribution is critical. Naive uniform sampling from raw logs inevitably introduces platform-specific biases; for instance, fashion queries might dominate on one platform while electronics dominate on another.

To mitigate this, we adopted an industry-stratified sampling strategy. We first categorized queries into distinct industry sectors and performed proportional sampling within each. Subsequently, we applied downsampling to head categories to flatten the distribution. This process yielded a balanced subset spanning \textbf{14 major industries}, ensuring that no single sector (e.g., Clothing or Cosmetics) constitutes more than 15\% of the total data, thereby guaranteeing a fair evaluation across diverse e-commerce domains.

\vspace{-0.2cm}
\subsection{Hard Subset}
\label{sec:hard}
We construct the Hard Subset to serve as a challenging testbed, pushing the boundaries of models' capabilities in distinguishing subtle relevance nuances. Based on the capability analysis in Section~\ref{sec:intro}, we construct the Hard Subset to specifically test Knowledge, Reasoning, and Multi-Attribute processing. We categorize challenging queries into three primary classes with specific sub-intents:

\textbf{1. Knowledge-Dependent Queries.}
These queries require external domain facts to bridge the gap between abstract user inputs and concrete items. We identify two specific sub-intents: \textbf{Domain Jargon (DJ)}, which involves specific model numbers or nicknames (e.g., ``13900'') that map to standard product categories; and \textbf{Entity Relationship (ER)}, which covers complex inter-product relations including both \textit{Alternatives} and \textit{Comparisons} (e.g., ``cheaper than iPhone'').

\textbf{2. Reasoning-Dependent Queries.}
These queries require logical inference to decode implicit user needs where the explicit text does not directly describe the target item. We distinguish three sub-intents: \textbf{Solution Seeking (SS)}, where users describe a usage scenario (e.g., ``bedroom noise reduction'' implies soundproofing materials); \textbf{Question Answering (QA)}, where users pose direct inquiries requiring functional understanding (e.g., ``how to clean suede shoes'' implies buying specialized cleaners); and \textbf{Negative Constraints (NC)}, which involve logical negation or exclusion terms (e.g., ``sugar-free'').

\textbf{3. Multi-Attribute Queries (MA).}
These queries represent explicit hard demands that simultaneously specify constraints across multiple dimensions in $\mathcal{D}$ and necessitate strict intersectional matching logic. To construct a robust dataset from massive logs, we implemented a rigorous three-step pipeline:

\textbf{Step 1: Hybrid Candidate Generation.}
We employed distinct mining strategies tailored to the linguistic characteristics of each intent:
(1) \textbf{Fixed-Pattern Mining (ER, QA, NC):} Since these intents exhibit stable linguistic structures, we defined expert rules and utilized regular expressions to extract candidates from logs.
(2) \textbf{Generative-Pattern Mining (SS):} As solution-seeking queries often follow diverse but extractable templates (e.g., ``gift for...'', ``outfit for...''), we first prompted Qwen3-235B to generate a comprehensive set of regex patterns, which were then used to mine the logs.
(3) \textbf{Long-tail Sampling (DJ):} We undersampled queries from the low-frequency spectrum of the search logs to capture domain-specific jargon.
(4) \textbf{Heuristic Filtering (MA):} We filtered raw candidates utilizing an internally developed Named Entity Recognition (NER) model. 
We specifically selected queries exhibiting high attribute density (e.g., 3<$\text{count}(\text{NER}(q))$<6) to maximize the presence of explicit constraints while mitigating noise. 
Detailed filtering protocols and model specifications are provided in Appendix~\ref{app:filtering}. 
Let the union of these raw candidates be $Q_{raw}$.

\textbf{Step 2: Tri-Model Intent Verification.}
The raw candidates inevitably contain noise. To confirm that the mined queries strictly match their target intents, we employed a Tri-Model Ensemble (Qwen3-235B, DeepSeek-R1, Llama3-70B). A query $q$ is retained only if the majority of models agree on its specific classification:
\begin{equation}
\label{equ:ensemble_voting}
Q_{verified} = \left\{ q \in Q_{raw} \;\middle|\; \sum_{k=1}^{3} \mathbb{I}(m_k(q) \in \mathcal{I}_{target}) \geq 2 \right\}
\end{equation}
where $\mathcal{I}_{target} = \{DJ, ER, SS, QA, NC, MA\}$ denotes the set of hard intents. Finally, to ensure the benchmark's reliability, the candidates in $Q_{verified}$ underwent a final expert review to validate intent classifications and assign ground-truth relevance labels, thereby eliminating any remaining model hallucinations.

\vspace{-0.1cm}
\subsection{Visually Salient Subset}
\label{sec:visual}
Fine-grained visual attributes (e.g., color, style, material) are imperative for accurate relevance assessment in e-commerce. While recent multimodal benchmarks like EcomMMMU~\cite{EcomMMMU} address general e-commerce tasks, they are not tailored for relevance judgment. Furthermore, applying their heavy model-based filtering directly to the massive scale of search logs is computationally prohibitive. To address these limitations, we devised a specialized, scalable multi-stage pipeline to identify query-item pairs where visual context is \textit{strictly necessary}.

\textbf{Taxonomy Grounding and Lexicon Expansion.}
Based on the attribute set $\mathcal{D}$ (Section~\ref{sec:3.2}), we identified nine core attributes requiring visual verification: IP, Material, Color, Set, Function, Style, Fashion Style, Season, and Audience. 
To align these abstract attributes with user search behaviors, we adopted a Generate-then-Filter strategy. 
Specifically, we prompted an LLM to generate a comprehensive synonym lexicon for each attribute, which was then used to retrieve an initial pool of candidate query-item pairs from the logs.

\textbf{Hierarchical Filtering Pipeline.}
To purify this candidate pool, we applied a rigorous two-stage filtering process:

\textbf{Stage 1: Coarse-grained Heuristic Filtering.} 
We first utilized a predefined blacklist to eliminate semantic false positives (e.g., filtering out ``milk powder'' incorrectly triggered by the color keyword ``pink''). This step effectively removes cases where keywords match superficially but diverge in semantic category, ensuring the initial quality of the candidate pool before fine-grained verification.

\textbf{Stage 2: Fine-grained VLM Ensemble Verification.} 
Acknowledging that static text-based rules may fail to capture nuanced visual-semantic misalignments, we employed a Tri-Model VLM Ensemble for final verification. We utilized three state-of-the-art VLMs—Qwen2.5-VL-72B, GLM-4.5v, and InternVL3-78B—to visually assess whether the judgment of the query-item pair strictly requires visual evidence. To ensure robustness and mitigate single-model hallucinations, we applied a majority voting mechanism. A query-item pair $x$ is retained in the final Visually Salient Subset $S_{VN}$ only if at least two models validate its visual necessity:
\begin{equation}
\label{equ:vlm_voting}
S_{VN} = \left\{ x \in S_{cand} \;\middle|\; \sum_{k=1}^{3} \mathbb{I}(V_k(x) = \text{essential}) \geq 2 \right\}
\end{equation}
where $S_{cand}$ denotes the filtered pairs from Stage 1, $V_k(\cdot)$ represents the binary verification output of the $k$-th VLM, and $\mathbb{I}(\cdot)$ is the indicator function. Consistent with the rigorous protocol of the Hard Subset, the final samples in $S_{VN}$ underwent expert human verification to confirm visual necessity and assign ground-truth relevance labels, thereby ensuring benchmark reliability.



\vspace{-0.1cm}
\subsection{Rule-Aware Benchmark}
\label{sec:rule-aware}
While explicit rules are vital for industrial relevance, public benchmarks typically lack such annotations. To bridge this gap and enable rigorous assessment of rule-following capabilities, we augment each RAIR sample with a checklist of governing rules, constructed via a model-assisted, human-verified pipeline.

\textbf{Model-based Pre-annotation.} We employed a reverse-engineering approach using Qwen3-235B-Instruct to generate initial candidates. To handle context limits, we first partitioned the full rule set $R_{\text{all}}$ into context-specific subsets $R'_{y,c}$ based on relevance label $y$ and industry $c$. For each sample $(Q, I)$, the LLM performed $k$ independent inferences. To ensure robustness and mitigate hallucination, we aggregated these results via majority voting:
\begin{equation}
R_{\text{cand}} = \left\{ r \in R'_{y,c} \mid \sum_{i=1}^{k} \mathbb{I}(r \in \mathcal{M}_i(Q, I, R'_{y,c})) \ge 2 \right\}
\end{equation}
where $\mathcal{M}_i$ denotes the rule set identified in the $i$-th inference.

\textbf{Human Verification.} Crucially, the LLM-generated set $R_{\text{cand}}$ serves only as a preliminary reference. Professional annotators reviewed each candidate rule, correcting false positives and recalling missed rules to produce the final ground-truth set $R_{(Q,I)}$. 
This human-in-the-loop process ensures the high fidelity required for a benchmark.

\vspace{-0.2cm}
\section{RAIR Benchmark}
\subsection{Data Statistics}
The RAIR benchmark spans 14 real-world e-commerce industries and comprises three distinct subsets totaling 48,949 samples: 32,123 in the General Subset, 10,931 in the Hard Subset, and 5,895 in the Visually Salient Subset. Table \ref{tab:query_freq} details the data volume across 6 challenging query intents within the Hard Subset, while the full industry distribution is provided in Appendix~\ref{app:B}.

Each entry contains the query, title, Category-Property-Value (CPV), Stock Keeping Unit (SKU), and image, alongside the manual L1--L4 label. 
We also provide the specific rule identifier used for judgment (Section~\ref{sec:rule-aware}). 
As shown in Table~\ref{tab:gt_distribution}, L4 (Perfect Match) and L2 (Explicit Mismatch) dominate the distribution. 
This imbalance is by design: L1 samples (noise) are trivial to discard and thus filtered during hard mining; conversely, L3 samples (partial matches) are naturally scarce boundary cases yet critical for distinguishing fine-grained semantics.

Table~\ref{tab:dataset_comparison} underscores RAIR's advantages over existing benchmarks, highlighting its superior feature diversity, complex data composition, and unique integration of explicit rules, all maintained at a substantial scale.

\begin{table}[htbp]
\centering
\vspace{-0.3cm}
\caption{Distribution of ground truth labels in RAIR}
\vspace{-0.3cm}
\label{tab:gt_distribution}
\resizebox{0.95\columnwidth}{!}{
\begin{tabular}{lccc}
\toprule
\textbf{Relevance} & \textbf{Ground Truth} & \textbf{Num} & \textbf{Frequency} \\
\midrule
\multirow{2}{*}{Irrelevant} & L1 & 2008 & 4.1\% \\
 & L2 & 10911 & 22.8\% \\
\midrule
\multirow{2}{*}{Relevant} & L3 & 2382 & 4.9\% \\
 & L4 & 33648 & 68.7\% \\
\bottomrule
\end{tabular}
}
\end{table}

\begin{table}[htbp]
  \centering
  \vspace{-0.3cm}
  \caption{Distribution of challenging query intentions within the Hard Subset.}
  \vspace{-0.3cm}
  \label{tab:query_freq}
  \resizebox{0.95\columnwidth}{!}{
  \begin{tabular}{llcc}
    \toprule
    \textbf{Category} & \textbf{Sub-Intent} & \textbf{Num} & \textbf{Freq.} \\
    \midrule
    \multirow{2}{*}{Knowledge-Dependent} 
      & Domain Jargon (DJ) & 2138 & 19.6\% \\
      & Entity Relationship (ER) & 213 & 1.9\% \\
    \cmidrule(lr){1-4} 
    
    \multirow{3}{*}{Reasoning-Dependent} 
      & Solution Seeking (SS) & 2192 & 20.1\% \\
      & Question Answering (QA) & 244 & 2.2\% \\
      & Negative Constraints (NC) & 1073 & 9.8\% \\
    \cmidrule(lr){1-4} 
    
    \multirow{1}{*}{Multi-Attribute} 
      & Multi-Attribute (MA) & 5071 & 46.4\% \\
    \bottomrule
  \end{tabular}
  }
\end{table}

\newcommand{\cmark}{\ding{51}}
\newcommand{\xmark}{\ding{55}}

\begin{table}[htbp]
    \centering
    \setlength{\tabcolsep}{5pt} 
    \vspace{-0.4cm}
    \caption{Comparison between \textbf{RAIR} and other e-commerce datasets. 
    ``\textbf{CR}'': Contains explicit Relevance tasks. 
    ``\textbf{Vis}'': Includes Visual modality. 
    ``\textbf{Rule}'': Provides explicit, complete assessment Rules. 
    ``\textbf{Hard}'': Explicitly constructs a Hard/Boundary subset. 
    ``\textbf{VS}'': Identifies a Visually Salient Subset where images are essential for judgment.
    }
    \vspace{-0.2cm}
    \label{tab:dataset_comparison}
    \resizebox{\columnwidth}{!}{%
    \begin{tabular}{lccccc}
        \toprule
        \textbf{Dataset} & \textbf{CR} & \textbf{Vis} & \textbf{Rule} & \textbf{Hard} & \textbf{VS} \\
        \midrule
        Amazon-M2~\cite{amazon}                 & \xmark & \xmark & \xmark & \xmark & \xmark \\
        EcomInstruct~\cite{eomgpt}              & \xmark & \xmark & \xmark & \xmark & \xmark \\
        Shopping Queries~\cite{shoppingqueries} & \cmark & \xmark & \xmark & \xmark & \xmark \\
        ECInstruct~\cite{Ecinstruct}            & \cmark & \xmark & \xmark & \xmark & \xmark \\
        Shopping MMLU~\cite{shopingmmmu}       & \cmark & \xmark & \xmark & \xmark & \xmark \\
        EcomMMMU~\cite{EcomMMMU}                & \cmark & \cmark & \xmark & \xmark & \cmark \\
        \midrule
        \textbf{RAIR (Ours)}                    & \textbf{\cmark} & \textbf{\cmark} & \textbf{\cmark} & \textbf{\cmark} & \textbf{\cmark} \\ 
        \bottomrule
    \end{tabular}%

    }
\end{table}


\vspace{-0.5cm}
\subsection{Evaluation Metrics}
\begingroup
\setlength{\abovedisplayskip}{3pt}
\setlength{\belowdisplayskip}{3pt}

We employ Acc@4 for fine-grained evaluation and Acc@2 for binary relevance, where $\{L1, L2\}$ are treated as irrelevant and $\{L3, L4\}$ as relevant. To mitigate label imbalance, we also utilize Macro-F1. The metrics are defined as:
\begin{align}
\label{eq:acc4}
\text{Acc@4} &= \frac{1}{N}\sum_{i=1}^{N}\mathbb{I}(\hat{y}_i = y_i) \\
\label{eq:acc2}
\text{Acc@2} &= \frac{1}{N} \sum_{i=1}^{N} \mathbb{I} \big( (\hat{y}_i, y_i \in \{L1, L2\}) \lor (\hat{y}_i, y_i \in \{L3, L4\}) \big) \\
\label{eq:macro_f1}
\text{Macro-F1} &= \frac{1}{C}\sum_{i=1}^{C} \frac{2 \cdot \text{Precision}_i \cdot \text{Recall}_i}{\text{Precision}_i + \text{Recall}_i}
\end{align}
\endgroup



\section{Experiments}



\vspace{-0.1cm}
\subsection{Setups}
We evaluate a diverse spectrum of models on RAIR, ranging from open-source LLMs (4B--235B parameters) to SOTA API models like GPT-5 and Gemini 2.5 Pro. For closed-source APIs, we evaluate on a randomly sampled subset of 1,000 instances due to cost constraints. \textit{Specialized models such as LREF~\cite{LREF} and ELLM~\cite{ELLM} are excluded due to unavailability.} For reproducibility, all evaluations use official parameters with greedy decoding. Given the significant class imbalance (L4: 68.7\%, Table~\ref{tab:gt_distribution}), a trivial majority baseline yields 68.7\% accuracy but only ~17.2\% F1. We therefore adopt Macro-F1 as the primary metric to robustly penalize such degenerate strategies.

\vspace{-0.2cm}
\subsection{Main Results}
\vspace{-0.1cm}

\begin{table}[!h]
\centering
\vspace{-0.3cm}
\caption{Results on the Visually Salient Subset (``w/o image'' denotes no image input), $^{\ast}$Used in Sec.~\ref{sec:visual}}
\vspace{-0.3cm}
\label{tab:visual_results}
\resizebox{0.95\columnwidth}{!}{
\begin{tabular}{lcccc}
\toprule
\multirow{2}{*}{\textbf{Model}} & \multirow{2}{*}{\textbf{Setting}} & \multicolumn{3}{c}{\textbf{Visually Salient Subset}}\\
\cmidrule(lr){3-5}
& & Acc@2 & Acc@4 & Macro-F1 \\
\midrule
\multicolumn{5}{l}{\textit{Open-source Instruct Models}} \\
\multirow{2}{*}{Qwen2.5-VL-7B-Instruct\cite{qwen2.5vl}} & w/o image & 0.513 & 0.271 & 0.217 \\
 & w/ image & 0.535 & 0.285 & 0.230 \\
\multirow{2}{*}{Qwen2.5-VL-32B-Instruct\cite{qwen2.5vl}} & w/o image & 0.551 & 0.391 & 0.250 \\
 & w/ image & 0.647 & 0.467 & 0.339 \\
\multirow{2}{*}{Qwen2.5-VL-72B-Instruct$^{\ast}$\cite{qwen2.5vl}} & w/o image & 0.535 & 0.394 & 0.248 \\
 & w/ image & 0.608 & 0.420 & 0.267 \\
 \multirow{2}{*}{GLM-4.5v\cite{GLM4.5v}} 
 & w/o image & 0.231 & 0.204 & 0.183 \\
  & w/ image & 0.295 & 0.278 & 0.223 \\
\midrule
\multicolumn{5}{l}{\textit{Closed-source Models}} \\
\multirow{2}{*}{Gemini 2.5 Pro\cite{gemini2.5}} & w/o image & 0.578 & 0.481 & 0.285 \\
 & w/ image & \underline{0.670} & \textbf{0.561} & \textbf{0.377} \\
\multirow{2}{*}{GPT-5} & w/o image & 0.618 & 0.414 & 0.298 \\
 & w/ image & \textbf{0.682} & \underline{0.508} & \underline{0.369} \\
\bottomrule
\end{tabular}
}
\end{table}

\begin{table*}[htbp]
\centering
\vspace{-0.3cm}
\caption{Main results on General and Hard Subsets. \textbf{Bold}: best; \underline{underlined}: second-best. $^{\ast}$Used in Sec.~\ref{sec:hard}.
}
\vspace{-0.3cm}
\label{tab:main_results}
\begin{tabular}{llcccccc}
\toprule
\multirow{2}{*}{\textbf{Model}} & \multirow{2}{*}{\textbf{Setting}} & \multicolumn{3}{c}{\textbf{General Set}} & \multicolumn{3}{c}{\textbf{Hard Subset}} \\
\cmidrule(lr){3-5} \cmidrule(lr){6-8}
& & Acc@2 & Acc@4 & Macro-F1 & Acc@2 & Acc@4 & Macro-F1 \\
\midrule
\multicolumn{8}{l}{\textit{Open-source Instruct Models}} \\
\multirow{2}{*}{Qwen3-4B-Instruct\cite{Qwen3}} & Base & 0.761 & 0.694 & 0.408 & 0.539 & 0.385 & 0.303 \\
 & w/ rule & 0.787 & 0.694 & 0.404 & 0.542 & 0.378 & 0.300 \\
\multirow{2}{*}{Qwen3-30B-Instruct\cite{Qwen3}} & Base & 0.766 & 0.707 & 0.409 & 0.586 & 0.442 & 0.343 \\
 & w/ rule & 0.763 & 0.683 & 0.391 & 0.587 & 0.431 & 0.333 \\
 \multirow{2}{*}{Qwen3-235B-Instruct$^{\ast}$\cite{Qwen3}} & Base & \underline{0.839} & 0.714 & 0.466 & 0.604 & 0.411 & 0.364 \\
 & w/ rule & 0.830 & 0.676 & 0.417 & 0.609 & 0.381 & 0.359 \\
 \multirow{2}{*}{Llama3.1-8B-Instruct\cite{llama3}} & Base & 0.786 & 0.529 & 0.261 & 0.542 & 0.314 & 0.227 \\
 & w/ rule & 0.788 & 0.416 & 0.224 & 0.525 & 0.248 & 0.200 \\
\multirow{2}{*}{Llama3.1-70B-Instruct\cite{llama3}} & Base & 0.781 & \textbf{0.721} & 0.395 & 0.536 & 0.431 & 0.304 \\
 & w/ rule & 0.774 & 0.668 & 0.369 & 0.547 & 0.378 & 0.295 \\
\midrule
\multicolumn{8}{l}{\textit{Open-source Thinking Models}} \\
\multirow{2}{*}{Qwen3-4B-Thinking\cite{Qwen3}}
& Base & 0.692	&0.614	&0.385& 0.524	&0.387	&0.311 \\
 & w/ rule & 0.805 & \underline{0.720} & 0.453 & 0.584 & 0.419 & 0.357 \\
\multirow{2}{*}{Qwen3-30B-Thinking\cite{Qwen3}} & Base & 0.739 & 0.671 & 0.424 & 0.521 & 0.398 & 0.306 \\
 & w/ rule & 0.784 & 0.718 & 0.457 & 0.582 & 0.448 & 0.362 \\
\multirow{2}{*}{Qwen3-235B-Thinking\cite{Qwen3}} & Base & 0.712	& 0.636 &	0.422 & 0.546	&0.412	&0.321 \\
 & w/ rule & 0.779 & 0.702 & \underline{0.470} & 0.585 & 0.451 & 0.367 \\
\midrule
\multicolumn{8}{l}{\textit{Closed-source Models}} \\
\multirow{2}{*}{Gemini 2.5 Pro\cite{gemini2.5}} & Base & 0.783 & 0.692 & 0.465 & 0.608 & \underline{0.466} & 0.389 \\
 & w/ rule & 0.795 & 0.701 & \textbf{0.483} & 0.627 & \textbf{0.481} & \underline{0.392} \\
\multirow{2}{*}{GPT-5} & Base & 0.781 & 0.654 & 0.434 & \underline{0.643} & 0.423 & 0.381 \\
 & w/ rule & \textbf{0.845} & 0.714 & 0.433 & \textbf{0.681} & 0.435 & \textbf{0.407} \\
\bottomrule
\end{tabular}
\end{table*}

Tables \ref{tab:visual_results} and \ref{tab:main_results} summarize the performance of open-source (sorted by size) and closed-source models, covering both \textit{Instruct} and \textit{Thinking} variants. We also report results using \textit{rule-guided inference} by injecting discriminative rules into the prompts. Two primary conclusions emerge:

\vspace{-0.1cm}
\subsubsection{RAIR is Challenging.} 
The benchmark poses significant difficulties even for top-tier models. Without rule guidance, GPT-5 achieves only 84.5\% Acc@2 and 43.3\% Macro-F1 on the General Subset. The difficulty escalates in the \textit{Hard} and \textit{Visually Salient} Subsets, where all models suffer a performance drop of at least 10 percentage points, underscoring RAIR's effectiveness as a stress test for robustness.

\vspace{-0.1cm}
\subsubsection{RAIR is Highly Discriminative.} 
RAIR effectively highlights performance disparities across model types and scales:

\textbf{(1) Open-source vs. Closed-source Gap.} While the leading open-source model (Qwen3-235B-Instruct) matches closed-source counterparts on the General Subset, a substantial gap persists in the Hard Subset (60.9\% vs. 68.1\% Acc@2 w/ rule). This confirms that closed-source models retain a distinct advantage in handling complex, long-tail scenarios.

\textbf{(2) Scale-Dependent Performance on Hard Cases.} While we generally observe that larger models perform better, scaling inversions occasionally occur on the General Subset. For instance, Qwen3-4B-Instruct (w/ rule) unexpectedly outperforms its 30B counterpart (78.7\% vs. 76.3\% in Acc@2). In stark contrast, the Hard Subset restores a strict positive correlation between parameter scale and performance (e.g., 54.2\% < 58.7\% < 60.9\% for Qwen3-4B/30B/235B). This distinction underscores that the General Subset may hit a saturation point, whereas the Hard Subset is essential for exposing the true capability limits and stratification of models.

\textbf{(3) Impact of Rule Integration.} Integrating explicit rules ($\sim$10k tokens) yielded negligible or negative gains for most Instruct models, whereas Thinking models exhibited significant improvements (e.g., GPT-5: $78.1\% \to 84.5\%$). 
This divergence suggests that Instruct models suffer from the ``lost-in-the-middle'' phenomenon, struggling to retrieve and apply guidelines from lengthy contexts. 
In contrast, Thinking models demonstrate superior long-context reasoning, enabling them to ground judgments in protocols. This confirms that \textit{rule-guided reasoning outperforms unconstrained speculation}, preventing capable models from over-complicating tasks.

\textbf{(4) Thinking vs. Non-Thinking Models.} Without rules, Thinking models underperform due to unconstrained hallucination; with rules, they approach Non-Thinking models. This implies that while Thinking models have potential, they require explicit constraints (like our rules) to ground their reasoning effectively.

\textbf{(5) Multimodal Capabilities.} As shown in Table \ref{tab:visual_results}, incorporating visual input consistently improves VLM performance on the Visually Salient Subset. Crucially, this gain scales with model capacity: Gemini-2.5-Pro shows substantial improvement (+9.2\% Acc@2), whereas Qwen2.5-VL-7B sees only marginal gains. This indicates that stronger models possess superior visual grounding and cross-modal integration abilities.

\vspace{-0.2cm}
\subsection{Comparative \& Fine-grained Analysis}
\vspace{-0.1cm}
\label{sec:6.3}
\subsubsection{Comparison with Existing Benchmarks.}
As shown in Table~\ref{tab:difficulty_comparison}, we benchmark RAIR-Hard against three existing datasets, strictly adhering to their original relevance scales and prompt configurations. Two critical observations underscore RAIR's value:

\textbf{High Difficulty and Prevention of Saturation.} As shown in Table~\ref{tab:difficulty_comparison}, while models on Shopping Queries and Shopping MMLU consistently exceed 0.80 accuracy, RAIR-Hard poses a substantial challenge. Even GPT-5 achieves only 0.681 Acc@2, with universal Macro-F1 scores below 0.41. This sharp performance drop underscores RAIR's effectiveness in exposing SOTA limitations within complex e-commerce contexts.

\begin{table}[htbp]
\centering
\vspace{-0.3cm}
\caption{Performance comparison across different e-commerce benchmarks.}
\vspace{-0.3cm}
\label{tab:difficulty_comparison}
\resizebox{0.95\columnwidth}{!}{
\begin{tabular}{llccc}
\toprule
\multirow{2}{*}{\textbf{Dataset}} & \multirow{2}{*}{\textbf{Model}} & \multicolumn{3}{c}{\textbf{Metrics}} \\
\cmidrule(lr){3-5}
 & & Acc@2 & Acc@4 & Macro-F1 \\
\midrule
\multirow{3}{*}{Shopping Queries\cite{shoppingqueries}} 
 & Qwen3-4B    & 0.877 & 0.656 & 0.408 \\
 & Qwen3-235B  & 0.813 & 0.662 & 0.469 \\
 & GPT-5       & 0.864 & 0.728 & 0.472 \\
\midrule
\multirow{3}{*}{EcomMMMU\cite{EcomMMMU}} 
 & Qwen3-4B    & 0.673 & 0.240 & 0.169 \\
 & Qwen3-235B  & 0.672 & 0.311 & 0.239 \\
 & GPT-5       & 0.703 & 0.342 & 0.248 \\
\midrule
\multirow{3}{*}{Shopping MMLU\cite{shopingmmmu}} 
 & Qwen3-4B    & 0.843 & 0.375 & 0.318 \\
 & Qwen3-235B  & 0.818 & 0.571 & 0.530 \\
 & GPT-5       & 0.839 & 0.548 & 0.492 \\
\midrule
\multirow{3}{*}{\textbf{RAIR-Hard (Ours)}} 
 & Qwen3-4B    & 0.542 & 0.378 & 0.300 \\
 & Qwen3-235B  & 0.609 & 0.381 & 0.359 \\
 & GPT-5       & 0.681 & 0.435 & 0.407 \\
\bottomrule
\end{tabular}
}
\end{table}

\textbf{Superior Discriminative Power.} Unlike existing benchmarks where performance gaps are often negligible or inverted (implying surface-level pattern fitting), RAIR exhibits strict stratification correlated with model scale: Qwen3-4B (0.542) < Qwen3-235B (0.609) < GPT-5 (0.681). This clear alignment confirms that RAIR measures rigorous reasoning rather than noise. By effectively distinguishing capabilities based on logical depth, RAIR serves as a reliable ``ruler'' for evaluating the advancements of future LLMs.

\vspace{-0.2cm}
\subsubsection{Sources of Difficulty in RAIR}
\begin{figure}[htbp]
    \centering
\includegraphics[width=1\linewidth]{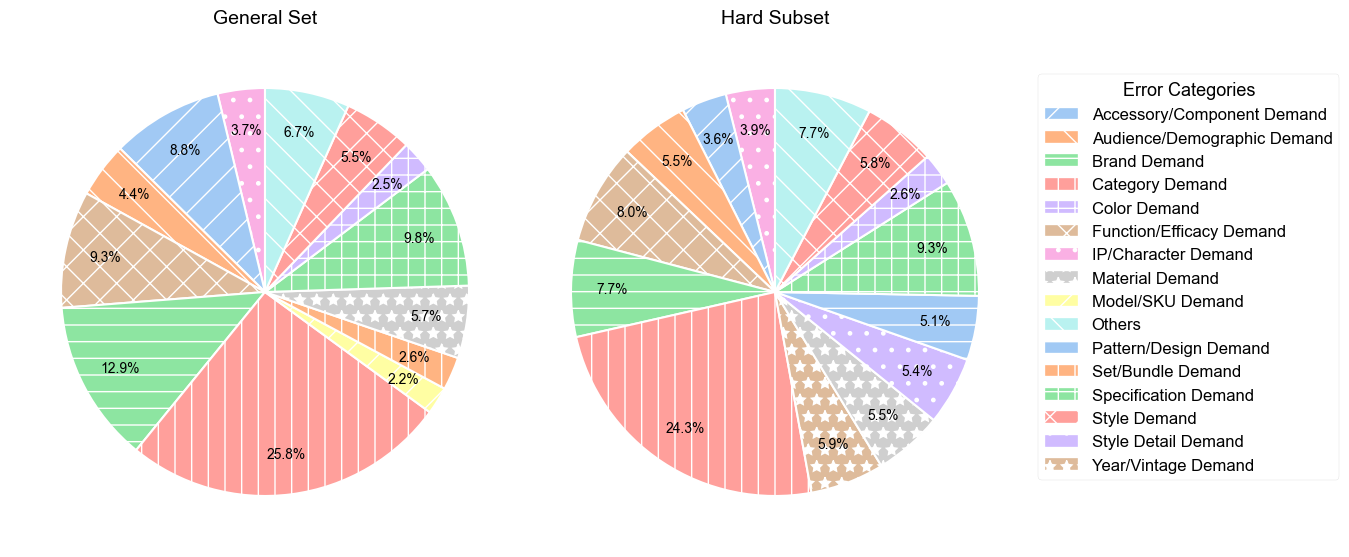}
    \vspace{-0.3cm}
    \caption{Distribution of error categories on the General set and the Hard Subset.}
        \vspace{-0.3cm}
    \label{fig:case_rule}
\end{figure}

To pinpoint the underlying sources of difficulty within RAIR, we analyzed the error distributions of the three top-performing models: Qwen3-235B-Instruct, GPT-5, and Gemini 2.5 Pro. By examining their failures across the General and Hard Subsets, we identified the specific rule violations that contribute most to the benchmark's challenge (Figure~\ref{fig:case_rule}).

\textbf{Fundamental Attributes dominate General Difficulty.} On the General Set, the primary challenge stems from basic entity alignment. \textit{Category Mismatches} account for 25.8\% of errors, followed by \textit{Specification} (12.9\%) and \textit{Brand} (9.8\%) inconsistencies. This indicates that RAIR's rigorous annotation pipeline ensures that even "easy" samples require precise entity verification, preventing models from relying on loose semantic associations.

\textbf{Abstract and Implicit Constraints drive Hard Difficulty.} In the Hard Subset, the difficulty profile shifts. While \textit{Category} issues persist (24.3\%), the relative difficulty of concrete specifications drops (12.9\% $\to$ 7.7\%), giving way to challenges involving abstract concepts and subjective user needs. This validates our hard-mining strategy: RAIR successfully increases difficulty not just by adding more parameters, but by introducing complex, implicit intents that demand higher-order reasoning. Further industry-specific analysis is provided in Appendix \ref{app:A}.

\vspace{-0.2cm}

\vspace{-0.4cm}
\section{Conclusion}
\vspace{-0.1cm}
We propose RAIR (Rule-Aware benchmark with Image for Relevance assessment), a comprehensive benchmark for evaluating e-commerce relevance in LLMs and VLMs. Distinctively, RAIR features a complete set of open-sourced judgment rules, with each instance explicitly annotated by rule ID. The benchmark comprises a General Subset for fundamental skills, alongside Hard and Visually Salient Subsets to probe performance upper bounds and multimodal proficiency, respectively. Extensive experiments demonstrate RAIR's challenging and comprehensive nature.

\bibliographystyle{ACM-Reference-Format}
\balance
\bibliography{main}
\newpage

\appendix

\begin{figure}
    \centering
    \includegraphics[width=1\linewidth]{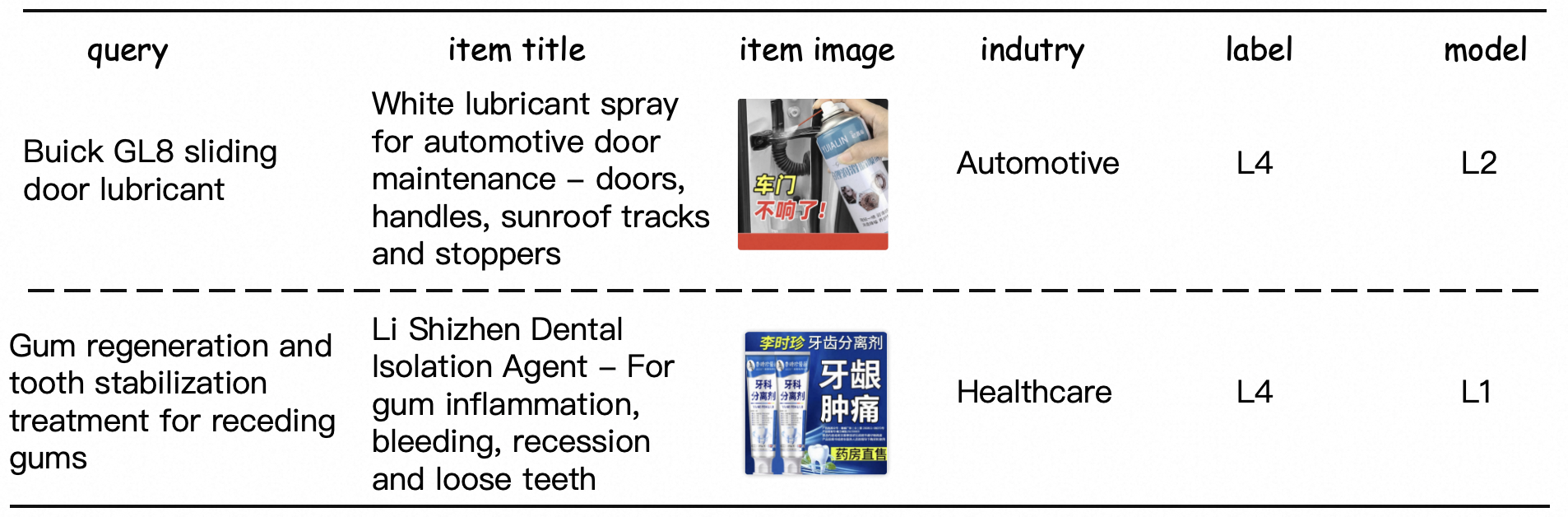}
    \caption{Industry hard case}
    \label{fig:hangye_case}
\end{figure}

\section{Industry-Specific Challenge Analysis}
\label{app:A}
To illustrate the domain-specific difficulties inherent in RAIR, we analyzed the performance of \texttt{Qwen3-235B} across the General Subset, identifying ``hard domains'' where accuracy lagged the overall mean by over 3 percentage points. As presented in Table \ref{tab:hangye_acc}, the Automotive and Healthcare sectors pose the most significant challenges, with performance gaps exceeding 7\% relative to the average. This disparity underscores the high barrier of domain knowledge required by RAIR. Figure \ref{fig:hangye_case} visualizes representative challenging samples from these sectors.

\textbf{Healthcare Sector Challenges.} The primary difficulty stems from the semantic gap between layperson queries and professional medical classifications. As shown in the bottom case of Figure \ref{fig:hangye_case}, accurate relevance judgment requires mapping a symptom-based query (e.g., ``gum regeneration'') to a specific regulatory product category (e.g., ``Dental Isolation Agent''). This demands that models benchmarked on RAIR possess deep specialized pharmaceutical knowledge to bridge the gap between colloquial symptoms and professional terminology.

\textbf{Automotive Sector Challenges.} Since e-commerce platforms predominantly sell parts and accessories, the challenge lies in rigorous compatibility verification. As illustrated in the top case of Figure \ref{fig:hangye_case} (a lubricant for a specific Buick GL8 model), the benchmark requires models to perform fine-grained attribute extraction to determine whether a general accessory is compatible with specific vehicle model constraints, testing their ability to handle strict intersectional logic.

\begin{table}[htbp]
  \centering  
  \caption{Industries significantly below the average level}
  \label{tab:hangye_acc}
  \begin{tabularx}{\columnwidth}{>{\centering\arraybackslash}X c}  
    \toprule
    Industry  & Acc@2  \\
    \midrule
    Average &    0.839 \\
    \midrule
    Food \& Fresh Products & 0.806 \\
    Fashion \& Apparel & 0.803 \\
    Industrial \& Agricultural & 0.778 \\
    Automotive & 0.760 \\
    Healthcare & 0.757 \\
    \bottomrule  
  \end{tabularx}
\end{table}

\begin{figure}
    \centering
    \includegraphics[width=1\linewidth]{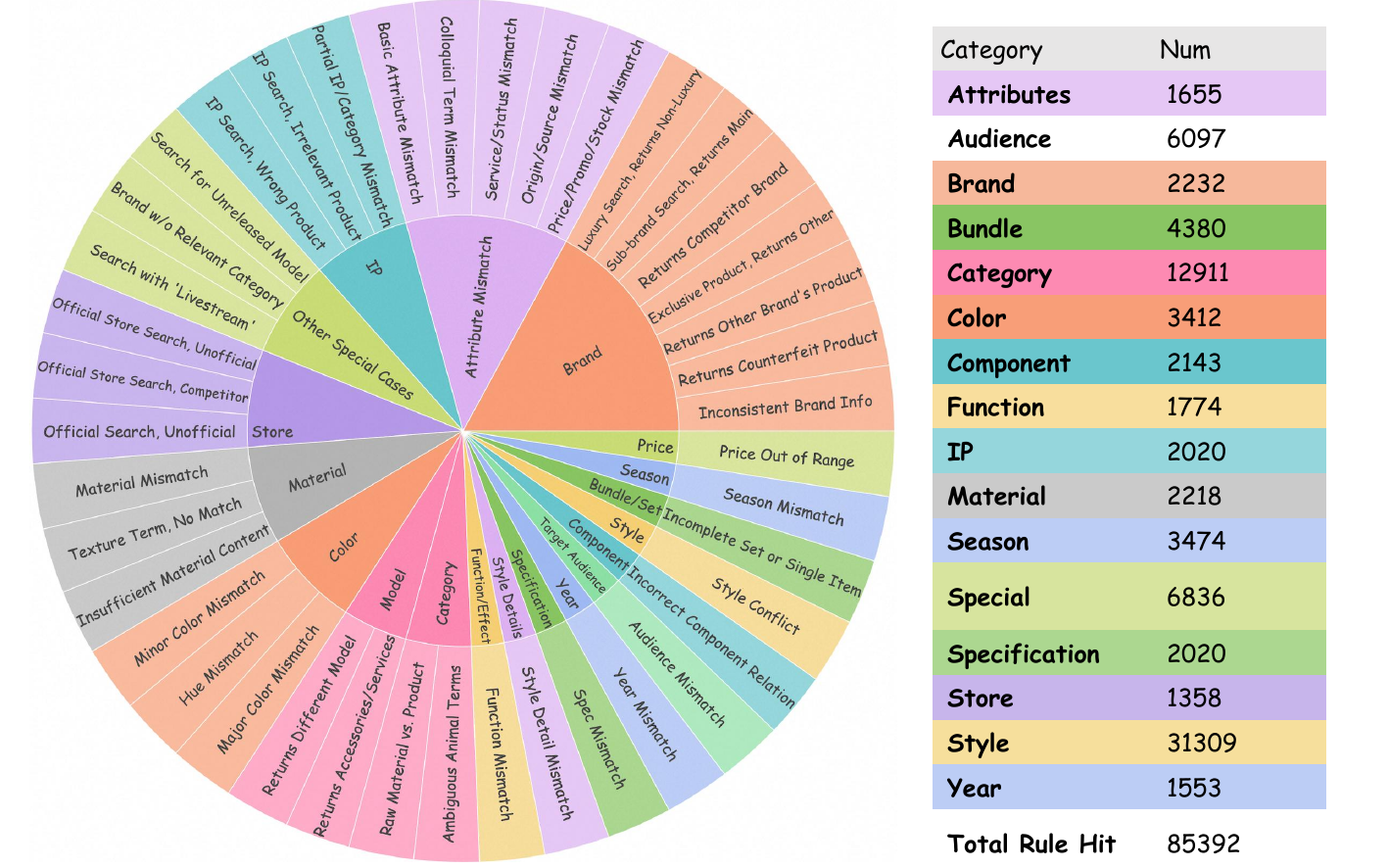}
    \caption{The Relevance Rule Framework of RAIR}
    \label{fig:RAIR_rule}
\end{figure}

\section{Supplementary Data Statistics}
\label{app:B}
\subsection{The Industry Distribution in General Subset}
\label{app:B.1}
\begin{figure}
    \centering
    \includegraphics[width=1\linewidth]{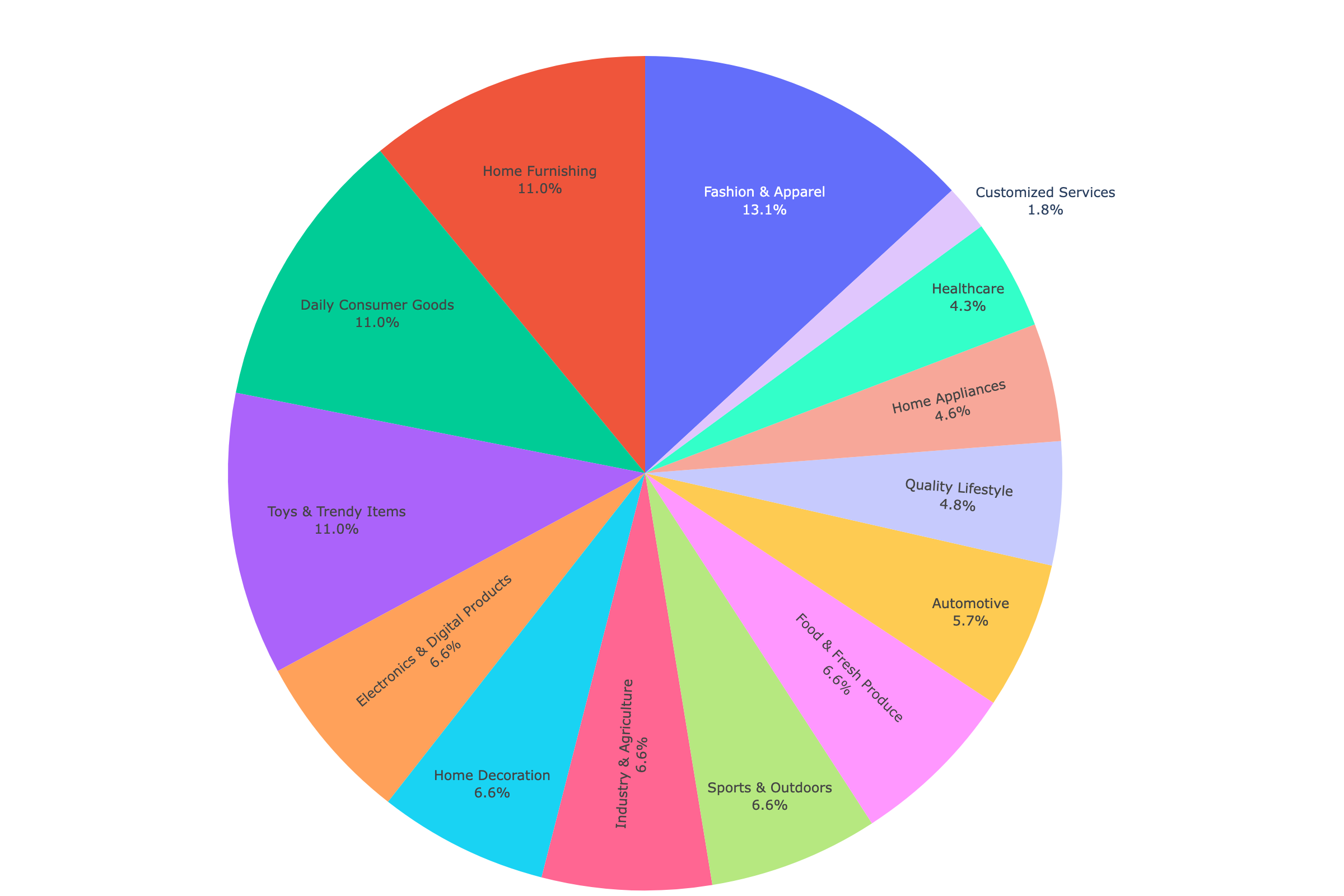}
    \caption{Industry Distribution in General Subset}
    \label{fig:hangye_distribution}
\end{figure}

Figure~\ref{fig:hangye_distribution} illustrates the industry distribution within the General Subset. A key characteristic is the \textbf{controlled balance} achieved through our stratified sampling strategy. Unlike raw search logs where fashion queries typically dominate, our subset caps the largest category, \textit{Fashion \& Apparel}, at just \textbf{13.1\%}. Major high-frequency sectors such as \textit{Home Furnishing}, \textit{Daily Consumer Goods}, and \textit{Toys \& Trendy Items} are strictly aligned at 11.0\% each to ensure uniformity. Furthermore, the dataset preserves significant diversity by covering specialized verticals like \textit{Automotive} (5.7\%) and \textit{Healthcare} (4.3\%), alongside long-tail needs like \textit{Customized Services} (1.8\%). This deliberate distribution prevents evaluation bias towards head categories, ensuring a comprehensive and fair assessment of model performance across diverse e-commerce domains.

\subsection{Distribution of Relevance Rules}
\label{sec:B.2}
To ensure objective and reproducible annotations, RAIR grounds every relevance judgment in a standardized framework. We established a comprehensive taxonomy comprising \textbf{16 distinct intent dimensions}, designed to capture the full spectrum of e-commerce user needs. The precise definition of each dimension is detailed in Table \ref{tab:rule_definitions}.

\begin{table}[H]
    \centering
    \caption{Definitions of the 16 intent dimensions constituting the RAIR relevance rule system.}
    \label{tab:rule_definitions}
    \resizebox{1.0\linewidth}{!}{
    \begin{tabular}{l|l}
        \toprule
        \textbf{Dimension} & \textbf{Definition} \\ 
        \midrule
        \textbf{Category} & The primary classification of the good (e.g., dress, smartphone). \\
        \textbf{Style} & Aesthetic style or visual design (e.g., vintage, Korean-style, thickened). \\
        \textbf{Special} & Special queries involving abstract needs, promotions, or new arrivals. \\
        \textbf{Audience} & Target demographics, including gender and age group (e.g., for kids, men). \\
        \textbf{Bundle} & Constraints on set completeness or quantity (e.g., suit vs. single jacket). \\
        \textbf{Season} & Applicable time or season of use (e.g., summer, winter thermal). \\
        \textbf{Color} & Visual color attributes (e.g., red, navy blue). \\
        \textbf{Brand} & Specific brand requirements (e.g., Nike, Apple). \\
        \textbf{Material} & Composition material of the product (e.g., cotton, leather). \\
        \textbf{Component} & Relationship between the main product and accessories/parts. \\
        \textbf{Specification} & Technical parameters (e.g., size, weight, capacity). \\
        \textbf{IP} & Intellectual Property rights or character associations (e.g., Disney, Marvel). \\
        \textbf{Function} & Specific efficacy or usage scenarios (e.g., whitening, gaming). \\
        \textbf{Attributes} & Other specific product properties not covered above (e.g., second-hand, origin). \\
        \textbf{Year} & Specific model year or vintage (e.g., 2023 version). \\
        \textbf{Store} & Constraints on the seller or channel (e.g., official flagship store). \\
        \bottomrule
    \end{tabular}
    }
\end{table}

Figure \ref{fig:RAIR_rule} visualizes this hierarchical rule taxonomy and the proportional distribution of rule applications within the dataset:

\begin{itemize}
    \item \textbf{Hierarchical Taxonomy (Sunburst Chart):} The chart on the left illustrates the fine-grained nature of our assessment protocols. Each core dimension branches into specific discriminative rules to handle edge cases. For instance, the \textit{Brand} dimension is not a monolithic label but is stratified into precise logic paths such as ``Competitor Brand,'' ``Sub-brand,'' or ``Counterfeit Product.'' This hierarchy ensures that models are evaluated on specific reasoning paths rather than vague semantic similarities.
    
    \item \textbf{Rule Distribution Analysis :} The table on the right highlights the distribution of rule activations across the dataset. \textbf{Visual and fundamental attributes}, particularly \textit{Style} and \textit{Category}, constitute the dominant proportion, reflecting their role as the primary decision factors in user search behavior. However, the dataset maintains a \textbf{substantial representation of complex, logic-heavy dimensions}, such as \textit{Audience}, \textit{Bundle/Set}, and \textit{Color}. This distribution ensures that while the benchmark captures common search needs, it retains a high density of intricate constraints required to rigorously test a model's fine-grained reasoning capabilities.
\end{itemize}

\section{Cross-lingual Applicability Analysis}
\label{app:cross_lingual}
\begin{table}[h]
\centering
\caption{Cross-lingual performance comparison on the General Set. ``CN'' denotes the original Chinese dataset, and ``EN'' denotes the GPT-5 translated English version. Despite a minor performance drop ($\Delta \approx 2\%$) due to translation noise, the results demonstrate the robustness of our logical rules across languages.}
\label{tab:cross_lingual_results}
\resizebox{0.95\columnwidth}{!}{
\begin{tabular}{lccccc}
\toprule
\multirow{2}{*}{\textbf{Model}} & \multirow{2}{*}{\textbf{Lang.}} & \multicolumn{3}{c}{\textbf{Metrics}} & \multirow{2}{*}{\textbf{Gap ($\Delta$)}} \\
\cmidrule(lr){3-5}
 & & Acc@2 & Acc@4 & F1 & \\
\midrule
\multirow{2}{*}{Qwen3-4B-Instruct} 
 & CN & 0.787 & 0.694 & 0.404 & - \\
 & EN & 0.765 & 0.678 & 0.391 & \textit{$\approx$ 2.2\%} \\ 
\midrule
\multirow{2}{*}{Qwen3-235B-Instruct} 
 & CN & 0.830 & 0.676 & 0.417 & - \\
 & EN & 0.811 & 0.659 & 0.406 & \textit{$\approx$ 1.9\%} \\ 
\bottomrule
\end{tabular}
}
\end{table}

Although RAIR is originally constructed on Chinese e-commerce data, a core premise of our benchmark is that the proposed relevance taxonomy and rule-based assessment logic are universal, reflecting fundamental human cognitive patterns in shopping rather than language-specific idiosyncrasies.

To verify this linguistic agnosticism, we conducted a cross-lingual transfer experiment using a rigorous \textbf{hybrid translation pipeline}. First, we utilized GPT-5 to translate the prompt templates, evaluation rules, and the entire General Set (including queries and item metadata) from Chinese into English. Subsequently, to ensure maximum semantic fidelity, all translated content underwent meticulous verification and refinement by bilingual domain experts. This human-in-the-loop process was crucial to correct potential hallucinations and ensure accurate mapping of culture-specific e-commerce terminology.

We evaluated two representative models, Qwen3-4B-Instruct and Qwen3-235B-Instruct, on this verified English version. As illustrated in Table~\ref{tab:cross_lingual_results}, the performance metrics on the English dataset exhibit a strong correlation with the original Chinese results, with performance variances restricted to approximately 2\%.

For instance, Qwen3-235B achieves an Acc@2 of 0.830 on the Chinese set and 0.811 on the English set. While a minor performance drop is observed—likely attributable to inevitable semantic shifts during translation—this structural consistency confirms that the constraints defined in RAIR are largely language-invariant. Models primarily rely on the logical reasoning structure provided by our rules rather than superficial linguistic cues, validating the robust cross-lingual applicability of the RAIR benchmark. The cross-lingual experiments primarily validate the universality of RAIR's logical rule structure. Meanwhile, we recognize that specific linguistic nuances—such as Chinese internet slang embedded in Domain Jargon queries—remain inherently non-transferable, underscoring the benchmark's inclusion of deep, culture-specific semantic challenges.

\section{Data Quality and Annotation Analysis}
\label{app:quality_control}

\subsection{Data Filtering and Pre-processing Details}
\label{app:filtering}
\subsubsection{NER Model Specifications}
To accurately identify queries with rich semantic constraints, we utilized an internally developed Named Entity Recognition (NER) model. This model is founded on the TBStar-13B architecture, a proprietary large language model pre-trained on massive e-commerce interaction logs, and underwent instruction fine-tuning specifically for e-commerce attribute extraction tasks.

To ensure the reliability of our filtering process, we evaluated the model on a held-out test set of manually annotated queries. As summarized in Table~\ref{tab:ner_perf_simple}, the model demonstrates robust performance with an overall Precision of 91.5\% and Recall of 90.8\%. This high fidelity ensures that the attribute counts used in our pipeline are statistically accurate, minimizing false positives in the complexity assessment.

\begin{table}[h]
\centering
\caption{Overall performance of the internal TBStar-13B based NER model on the held-out test set.}
\label{tab:ner_perf_simple}
\resizebox{0.6\linewidth}{!}{
\begin{tabular}{lc}
\toprule
\textbf{Metric} & \textbf{Performance Score} \\
\midrule
Overall Precision & 91.5\% \\
Overall Recall    & 90.8\% \\
Overall F1-Score  & 91.1\% \\
\bottomrule
\end{tabular}
}
\end{table}

\subsubsection{Threshold Calibration for Attribute Density}
To determine the optimal complexity threshold, we analyzed a random sample of 1,000 queries stratified by their NER entity counts ($N$). We evaluated these samples using two key metrics:
\begin{itemize}
    \item \textbf{Semantic Density (\%):} The proportion of queries containing sufficient multi-dimensional constraints to be categorized as "Hard" evaluation targets.
    \item \textbf{Naturalness Rate (\%):} The proportion of queries representing coherent, human-like search intent, distinguishing natural search behavior from raw product titles or SEO keyword stacking.
\end{itemize}

As evidenced in Table~\ref{tab:threshold_analysis}, the data reveals a critical quality boundary. 
At the lower end ($N=3$), the Semantic Density is insufficient (18.5\%), indicating that such queries are too simplistic for robust evaluation.
Crucially, at the upper end, we observe a precipitous drop in naturalness starting at $N=6$. While $N=4$ and $N=5$ maintain relatively high Naturalness Rates (81.5\% and 69.3\% respectively) alongside high Semantic Density, the Naturalness Rate collapses to 29.2\% at $N=6$. This sharp decline indicates that queries with 6 or more entities predominantly degenerate into unnatural keyword stuffing or incoherent lists.

Consequently, we identified the interval $N \in \{4, 5\}$ as the optimal "sweet spot," where the benchmark maximizes complexity without compromising linguistic validity. Based on these findings, we established the filtering constraint $3 < \text{count}(\text{NER}(q)) < 6$.

\begin{table}[h]
\centering
\caption{Analysis of query quality metrics across different NER entity counts ($N$).}
\label{tab:threshold_analysis}
\resizebox{0.85\linewidth}{!}{
\begin{tabular}{cccc}
\toprule
\textbf{\begin{tabular}[c]{@{}c@{}}NER Entity\\ Count ($N$)\end{tabular}} & 
\textbf{\begin{tabular}[c]{@{}c@{}}Sample\\ Proportion\end{tabular}} & 
\textbf{\begin{tabular}[c]{@{}c@{}}Semantic\\ Density (\%)\end{tabular}} & 
\textbf{\begin{tabular}[c]{@{}c@{}}Naturalness\\ Rate (\%)\end{tabular}} \\ 
\midrule
3 & 42.6\% & 18.5 & 91.2 \\
4 & 28.4\% & 76.8 & 81.5 \\
5 & 14.7\% & 82.4 & 69.3 \\
6 & 8.3\% & 88.1 &  29.2 \\
7 & 4.1\% & 91.5 &  11.7\\
8 & 1.9\% & 94.2 & 3.2 \\ 
\bottomrule
\end{tabular}
}
\end{table}

\subsection{Performance of LLM-Assisted Construction}
To balance scalability with precision, we leverage Large Language Models (LLMs) to assist in three specific stages of dataset construction. We adopted distinct optimization strategies for each stage based on their specific roles:

\begin{itemize}
    \item \textbf{Hard Query Filtering:} We prioritized \textbf{Recall} to ensure comprehensiveness. The relatively high recall (82.1\%) ensures that we capture a wide spectrum of potential boundary cases, minimizing the risk of missing valuable hard samples. Although precision is lower (71.4\%), false positives are easily removed during human verification.
    \item \textbf{Visual Necessity Filtering:} We prioritized \textbf{Precision} to enhance efficiency. Given the massive scale of e-commerce logs, our goal was to rapidly extract a high-quality subset of visual-dependent samples. The higher precision (79.6\%) allows annotators to focus on valid candidates, significantly reducing wasted effort on non-visual samples.
    \item \textbf{Rule Extraction:} This stage required a balance of both metrics to ensure accurate rule grounding, achieving $>$75\% in both precision and recall.
\end{itemize}

To validate these strategies, we conducted a pilot study on a sampled subset (N=1,000) for each stage. As shown in Table~\ref{tab:llm_performance}, the performance aligns well with our construction objectives. \textbf{Crucially, the LLM serves only as a preliminary candidate generator.} All outputs underwent mandatory human verification, ensuring that the strategic trade-offs in the LLM stage did not compromise the final dataset quality.

\begin{figure}[!t]
    \centering
    \begin{tcolorbox}[
        colback=gray!5,
        colframe=black,
        coltitle=white,
        fonttitle=\bfseries,
        title=Prompt: E-commerce Relevance Assessment,
        sharp corners,
        boxrule=0.8pt,
        left=2pt, right=2pt, top=2pt, bottom=2pt
    ]
    \small
    You are an expert in e-commerce relevance analysis. Your task is to assess the relevance between a user \textbf{Query} and an \textbf{item} based on provided information.

    \vspace{0.3em}
    \textbf{Key Attributes:} Pay attention to Category, Brand, Style Details, Material, Target Audience, Season, Model ID, Specifications, Color, Function, Accessories, Set/Single, Price, Year, Special Attributes, and IP image.

    \textbf{Instructions:}
    \begin{itemize}
        \setlength{\itemsep}{0pt}
        \item \textbf{Synthesize Information:} Combine insights from the item \textbf{Title}, \textbf{CPV} (Color-Pattern-Version), and \textbf{SKU}.
        \item \textbf{Chain of Thought (CoT):} You must enclose your reasoning process within \texttt{<think>} and \texttt{</think>} tags before outputting the final result.
        \item \textbf{Output Format:} For L2 cases, specify the mismatch type in brackets, e.g., \texttt{[L2-Style Mismatch]}.
    \end{itemize}

    \textbf{Relevance Levels:}
    \begin{itemize}
        \setlength{\itemsep}{0pt}
        \item \textbf{L1 (Irrelevant):} Complete category mismatch with no association.
        \item \textbf{L2 (Partially Irrelevant):} Category mismatch but related; or Category matches but key attributes (e.g., Brand, Spec, Gender) fail.
        \item \textbf{L3 (Closely Relevant):} Proximate category/attributes but lacks full intent alignment; or contains minor attribute conflicts.
        \item \textbf{L4 (Perfectly Relevant):} Completely satisfies the query intent.
    \end{itemize}

    \vspace{0.3em}
    \textbf{Input:} \\
    User Query: \{query\} \\
    Item Title: \{title\} \\
    Item Shop: \{shop\_name\} \\
    Item CPV: \{cpv\_info\} \\
    Item SKU: \{sku\_info\} 
    \end{tcolorbox}

    \vspace{-0.2cm}
    \caption{The prompt template used for relevance assessment. We explicitly instruct the model to treat high-CTR items as reference-only to mitigate popularity bias.} 
    \label{fig:sft_prompt} 
\end{figure}

\begin{figure}
    \centering
    \includegraphics[width=1\linewidth]{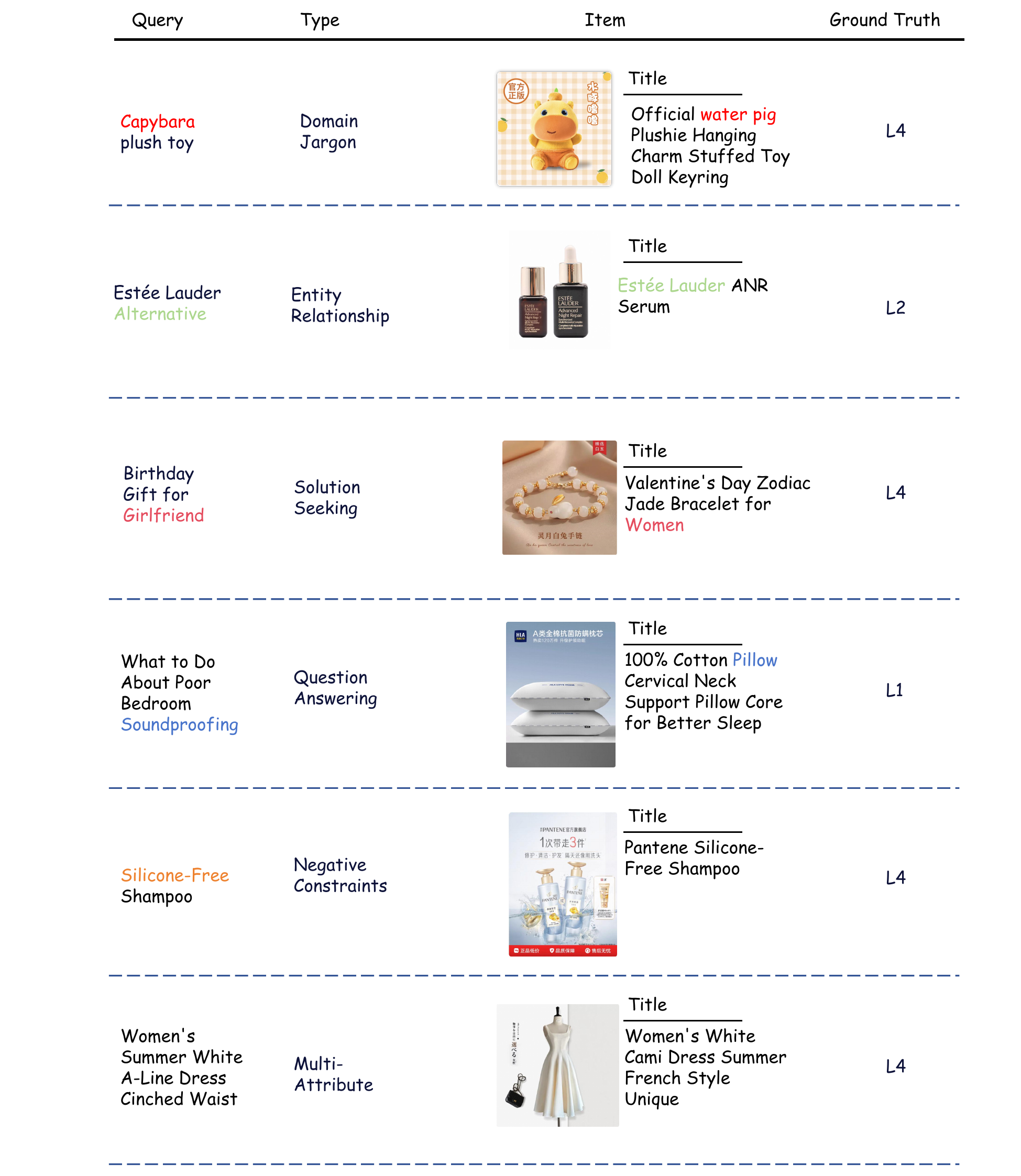}
    \caption{Representative examples from the RAIR Hard Subset across three difficulty categories: \textbf{Knowledge-Dependent} (Domain Jargon, Entity Relationship), \textbf{Reasoning-Dependent} (Solution Seeking, QA, Negative Constraints), and \textbf{Multi-Attribute} intents. These cases highlight the necessity for external knowledge, logical deduction, and strict constraint verification beyond simple keyword matching.}
    \label{fig:case_hard}
\end{figure}

\begin{table}[h]
\centering
\caption{Performance validation of LLMs in the data construction pipeline. The metrics reflect strategic trade-offs: \textbf{Hard Query Filtering} favors Recall for coverage, while \textbf{Visual Necessity Filtering} favors Precision for annotation efficiency.}
\label{tab:llm_performance}
\resizebox{0.95\columnwidth}{!}{
\begin{tabular}{lccc}
\toprule
\textbf{Construction Stage} & \textbf{Precision} & \textbf{Recall} & \textbf{Human Verified?} \\
\midrule
Hard Query Filtering       & 71.4\% & 82.1\% & \cmark \\
Visual Necessity Filtering & 79.6\% & 71.9\% & \cmark \\
Rule Extraction            & 75.2\% & 86.3\% & \cmark \\
\bottomrule
\end{tabular}
}
\end{table}

\subsection{Formulation of the Relevance Rule System}
To minimize subjective bias and ensure the generalization of our evaluation criteria across the 16 defined attributes, we implemented a rigorous, multi-stage rule formulation protocol.

\textbf{1. Expert Qualifications.}
The rule system was developed by a team of senior domain experts. To ensure high-quality standards, every participant was required to have a minimum of \textbf{five years of operational experience} in e-commerce relevance annotation. Their deep understanding of user search intent and query-item matching logic served as the foundation of this benchmark.

\textbf{2. Iterative Formulation Workflow.}
We adopted a collaborative and adversarial iterative process to refine the rules for each attribute:
\begin{itemize}
    \item \textbf{Group-based Drafting:} The experts were divided into small working groups, with \textbf{four members} per group. Each group was assigned specific attributes (e.g., Brand, Category) to draft the initial mapping rules from attribute satisfaction to relevance levels (L1--L4).
    \item \textbf{Intra-group Consensus:} Within each group, members debated boundary cases until a unanimous consensus was reached on the drafted rules.
    \item \textbf{Adversarial Cross-Validation:} Once a group finalized their draft, the rules were exchanged with another group for blind review. The reviewing group acted as ``adversaries,'' specifically looking for ambiguities, edge cases, or logical loopholes in the definitions.
    \item \textbf{Three-Round Iteration:} This cycle of drafting, consensus-building, and cross-examination was repeated for \textbf{three rounds}. The rules were finalized only after stabilizing across all teams, ensuring that the final protocol is both robust and reproducible.
\end{itemize}

\subsection{Annotation Pipeline and Quality Assurance}
To guarantee the gold-standard quality of RAIR, we implemented a rigorous, four-stage cascading quality control pipeline: \textbf{Initial Annotation $\rightarrow$ 1st Round QA $\rightarrow$ 2nd Round QA $\rightarrow$ Final Audit}.

Strict gating mechanisms were enforced at every transition. A batch was allowed to proceed to the next stage only if the accuracy in the current stage exceeded \textbf{95\%}. Batches failing this threshold were returned for complete re-annotation. This iterative refinement ensures that errors are intercepted early. As demonstrated in Table~\ref{tab:annotation_qa}, the final audit on the released dataset reveals an accuracy of \textbf{96.3\%}, attesting to the reliability of our benchmarks.

Furthermore, to quantify label consistency, we calculated the Inter-Annotator Agreement (IAA) across the three subsets. As shown in Table~\ref{tab:iaa}, the General set achieves a high agreement rate of 92.4\%, confirming the clarity of our rule system. Even for the more challenging Hard and Visually Salient Subsets, agreement remains robust at 89.1\% and 87.7\% respectively, indicating that our standardized protocol effectively aligns expert judgments even on complex and ambiguous cases.

\begin{table}[h]
\centering
\caption{Quality control statistics across the annotation pipeline. The strict threshold at each stage ensures a high-quality final deliverable.}
\label{tab:annotation_qa}
\resizebox{0.95\columnwidth}{!}{
\begin{tabular}{llcc}
\toprule
\textbf{Stage} & \textbf{Role} & \textbf{Threshold} & \textbf{Pass Rate} \\
\midrule
Phase 1 & Initial Annotation & - & - \\
Phase 2 & 1st Round QA (Peer Review) & $>$ 95\% & 95.4\% \\
Phase 3 & 2nd Round QA (Expert Review) & $>$ 95\% & 96.1\% \\
\midrule
\textbf{Final} & \textbf{Final Audit (Random Sampling)} & \textbf{-} & \textbf{96.3\%} \\
\bottomrule
\end{tabular}
}
\end{table}

\begin{table}[h]
    \centering
    \caption{Inter-Annotator Agreement (IAA) across RAIR subsets.}
    \label{tab:iaa}
    \begin{tabular}{lc}
    \toprule
    \textbf{Subset} & \textbf{Agreement Rate (\%)} \\
    \midrule
    General Subset & 92.4 \\
    Hard Subset & 89.1 \\
    Visually Salient Subset & 87.7 \\
    \bottomrule
    \end{tabular}
\end{table}

\begin{figure}[!t]
    \centering
    \begin{tcolorbox}[
        colback=gray!5,
        colframe=black,
        coltitle=white,
        fonttitle=\bfseries,
        title=Prompt: Hard Query Intent Classification,
        sharp corners,
        boxrule=0.8pt,
        left=2pt, right=2pt, top=2pt, bottom=2pt
    ]
    \small
    You are a professional e-commerce analyst. Your task is to analyze the user \textbf{Query} and classify its intent into one of the following \textbf{6 categories}.

    \vspace{0.3em}
    \textbf{Category Definitions:}
    \begin{enumerate}
        \item \textbf{Domain Jargon (DJ):} Queries containing specialized model numbers, nicknames, or industry terms requiring external knowledge to map to a product category (e.g., ``13900'' $\to$ CPU).
        \item \textbf{Entity Relationship (ER):} Queries seeking alternatives, comparisons, or specific relations between products (e.g., ``SK-II substitute'', ``cheaper than iPhone'').
        \item \textbf{Solution Seeking (SS):} Queries describing a usage scenario or problem without specifying a concrete product. (e.g., ``gift for girlfriend'', ``bedroom noise reduction''). \textit{Note: Specific requests like ``1.9m plastic bag'' are NOT SS.}
        \item \textbf{Question Answering (QA):} Direct inquiries about product functions or knowledge (e.g., ``Is Redmi a sub-brand of Xiaomi?'', ``how to clean suede'').
        \item \textbf{Negative Constraints (NC):} Queries explicitly excluding certain features or ingredients (e.g., ``silicone-free shampoo'', ``non-stick pan'').
        \item \textbf{Multi-Attribute (MA):} Queries explicitly specifying hard constraints across multiple dimensions (e.g., ``red nike running shoes size 42'').
        \item \textbf{Other:} Queries that do not fit into the above categories.
    \end{enumerate}

    \textbf{Instructions:}
    \begin{itemize}
        \setlength{\itemsep}{0pt}
        \item \textbf{Analyze Carefully:} First, identify if the query requires external knowledge (DJ/ER), logical reasoning (SS/QA/NC), or strict multi-condition matching (MA).
        \item \textbf{Strict Classification:} The result must be exactly one of the above 7 options.
        \item \textbf{Output Format:} First output your reasoning process in \texttt{<think>} tags, then provide the final category wrapped in \texttt{<answer>} tags.
    \end{itemize}

    \vspace{0.3em}
    \textbf{Input:} \\
    User Query: \{query\}
    \end{tcolorbox}

    \vspace{-0.2cm}
    \caption{The prompt template for categorizing hard query intents. It guides the model to map user inputs into specific complexity types such as Domain Jargon or Solution Seeking.}
    \label{fig:hard_query_prompt}
\end{figure}

\begin{figure}[!t]
    \centering
    
    \begin{tcolorbox}[
        colback=gray!5,
        colframe=black,
        coltitle=white,
        fonttitle=\bfseries,
        title=Prompt: Visual Necessity Assessment,
        sharp corners,
        boxrule=0.8pt,
        left=2pt, right=2pt, top=2pt, bottom=2pt
    ]
    \small
    You are a top-tier expert in e-commerce relevance analysis. Based on the provided item text and image, determine whether \textbf{visual information is a strictly necessary condition} for resolving the relevance judgment.

    \vspace{0.3em}
    \textbf{Metric: Visual Necessity} \\
    Evaluate the text and image comprehensively. Assign a value of \textbf{1} or \textbf{0}:
    \begin{itemize}
        \setlength{\itemsep}{0pt}
        \item \textbf{1 (Yes):} Must meet \textbf{BOTH} conditions:
        \begin{enumerate}
            \item \textbf{Text Insufficiency:} The title, shop name, and text details alone are insufficient to confirm relevance (due to ambiguity or missing key attributes).
            \item \textbf{Visual Solution:} The image provides critical visual features not mentioned or unclear in the text (e.g., specific style details, silhouette, actual color, spatial structure) that enable a definitive judgment.
        \end{enumerate}
        \item \textbf{0 (No):}
        \begin{itemize}
            \item The text information is already sufficient to judge relevance (whether relevant or irrelevant);
            \item OR, although the text is insufficient, the image also fails to provide the critical evidence needed.
        \end{itemize}
    \end{itemize}

    \textbf{Instructions:}
    \begin{itemize}
        \setlength{\itemsep}{0pt}
        \item \textbf{Internal Reasoning:} First, deeply analyze the core intent of the query, contrast it with the text coverage, and verify if the image provides decisive incremental information.
        \item \textbf{Strict Output:} You must AND ONLY need to provide a single number list in the format \texttt{\textbackslash boxed\{[X]\}}, where X is 0 or 1. Do not output any other content.
    \end{itemize}

    \vspace{0.3em}
    \textbf{Input:} \\
    User Query: \{query\} \\
    Item Title: \{title\} \\
    Item Shop: \{shop\_name\} \\
    Item CPV: \{cpv\_info\} \\
    Item SKU: \{sku\_info\} \\
    Item Image: \texttt{<image>}
    \end{tcolorbox}

    \vspace{-0.2cm}
    \caption{The prompt template for Visual Necessity Assessment. It instructs the model to determine if visual input is indispensable for addressing textual ambiguity.}
    \label{fig:visual_prompt}
\end{figure}

\section{Prompt Design and Reproducibility}
To ensure the reproducibility of RAIR and standardize the evaluation protocol, we explicitly provide the prompt templates tailored for the three critical stages of our pipeline: \textbf{Relevance Assessment}, \textbf{Hard Query Mining}, and \textbf{Visual Necessity Assessment}. These templates, visualized in Figures \ref{fig:sft_prompt}, \ref{fig:hard_query_prompt}, and \ref{fig:visual_prompt}, were iteratively refined to elicit optimal reasoning capabilities from LLMs.

\begin{itemize}
\item \textbf{Relevance Assessment (Figure \ref{fig:sft_prompt}):} This prompt instructs models to adhere to our 16-dimension intent taxonomy with a \textit{Chain-of-Thought} mechanism, requiring explicit reasoning about attribute mismatches before assigning L1--L4 labels. We also instruct models to treat high-CTR items as auxiliary context to mitigate popularity bias.

\item \textbf{Hard Query Mining (Figure \ref{fig:hard_query_prompt}):} This prompt categorizes queries by cognitive complexity, guiding models to distinguish between \textit{Knowledge-Dependent} intents (e.g., Domain Jargon) and \textit{Reasoning-Dependent} needs (e.g., Solution Seeking) to filter trivial samples.

\item \textbf{Visual Necessity Assessment (Figure \ref{fig:visual_prompt}):} This prompt performs a counterfactual check: determining whether visual information provides decisive value over text alone, ensuring evaluation targets cases where multimodal integration is strictly necessary.

\end{itemize}

\section{Case study}
Figure \ref{fig:case_hard} presents representative examples from the RAIR Hard Subset, illustrating the specific challenges across our three defined difficulty categories. These cases demonstrate why simple semantic matching is insufficient and where rigorous reasoning is required.

\newpage
\textbf{1. Knowledge-Dependent Queries (DJ \& ER).}
These cases test the model's ability to map abstract terms to specific entities using external facts.
\begin{itemize}
    \item \textbf{Domain Jargon (DJ):} In the first case, the user queries for a ``\textit{Capybara plush toy}.'' The relevant item is titled ``\textit{Official Water Pig...}'' (a common nickname for Capybara in specific cultural contexts). A model lacking this specific external knowledge graph would fail to link the nickname ``Water Pig'' to the scientific name ``Capybara,'' erroneously predicting a mismatch.
    \item \textbf{Entity Relationship (ER):} In the second case, the user explicitly seeks an ``\textit{Estée Lauder Alternative}'' (a substitute). The retrieved item is the \textit{original} Estée Lauder serum. While semantically highly similar, the relational intent is diametrically opposed (Substitute vs. Target). The model must understand the logical relationship implied by ``Alternative'' to correctly identify this as a mismatch (L2), rather than being misled by keyword overlap.
\end{itemize}

\textbf{2. Reasoning-Dependent Queries (SS, QA, \& NC).}
These cases require logical deduction to decode implicit needs or handle complex logical operators.
\begin{itemize}
    \item \textbf{Solution Seeking (SS):} The query ``\textit{Birthday Gift for Girlfriend}'' contains no concrete product keywords. The model must infer a usage scenario-to-product mapping, recognizing that a ``\textit{Jade Bracelet}'' functions as a valid gift option (L4), despite the lack of textual overlap.
    \item \textbf{Question Answering (QA):} For the query ``\textit{What to do about poor bedroom soundproofing},'' the candidate item is a ``\textit{Cervical Pillow}.'' While both relate to the broad topic of sleep quality, the pillow does not functionally address the user's specific question about noise isolation. The model must distinguish between a topical association and a functional solution to correctly label this as Irrelevant (L1).
    \item \textbf{Negative Constraints (NC):} The query ``\textit{Silicone-Free Shampoo}'' imposes a logical negation. The model must perform a strict exclusion check, verifying that the item attributes explicitly satisfy the ``non-existence'' of silicone, preventing hallucinated matches.
\end{itemize}

\textbf{3. Multi-Attribute Queries (MA).}
The final case, ``\textit{Women's Summer White A-Line Dress...},'' represents the challenge of strict intersectional logic. The query imposes simultaneous constraints across four dimensions: Gender, Season, Color, and Style. The model must verify that the item satisfies \textit{all} conditions conjunctively; missing even one attribute (e.g., if the dress were a straight cut instead of A-line) would degrade the relevance level, testing the model's precision in instruction following.

\section{Limitations and Potential Biases}
\label{app:limitations}

While RAIR establishes a rigorous framework for relevance assessment, we acknowledge inherent limitations stemming from the model-assisted construction pipeline. Specifically, we address the potential for \textbf{implicit bias} and \textbf{circular dependency} regarding the Qwen model family, which served dual roles in both data construction and final evaluation.

\subsection{Model Overlap and Data Leakage Risks}
As detailed in Section~\ref{sec:hard} and \ref{sec:visual}, high-performing LLMs were employed to scale the data curation process. We identify two specific instances where model overlap typically raises concerns about data leakage:

\begin{itemize}
    \item \textbf{Overlap in Hard Subset (Textual):} The \textit{Tri-Model Intent Verification} (Eq.~2) utilized \textbf{Qwen3-235B-Instruct} alongside \textbf{Llama3-70B} and \textbf{DeepSeek-R1} to filter candidate queries via majority voting. In our experiments (Table~\ref{tab:main_results}), the Qwen3 family (including the 235B variant) is evaluated on this very subset.
    
    \item \textbf{Overlap in Visual Subset (Multimodal):} The \textit{VLM Ensemble} (Eq.~3) for the Visually Salient subset employed \textbf{Qwen2.5-VL-72B} as a verifier. Subsequently, this model appears as an evaluation target in Table~\ref{tab:visual_results}.
\end{itemize}

This overlap theoretically introduces a "home-court advantage," where the evaluation data might implicitly align with the internal probability distributions or biases of the Qwen models, potentially inflating their scores.

\subsection{Mitigation Factors and Task Orthogonality}
Despite these overlaps, we argue that the benchmark retains its validity and discriminative power due to two key mitigating factors:

\begin{enumerate}
    \item \textbf{Task Orthogonality:} There is a fundamental cognitive distinction between the \textit{construction tasks} and the \textit{evaluation task}. 
    During construction, the models performed **Meta-Cognitive Verification**—assessing \textit{whether} a query is logically complex (Hard Subset) or necessitates visual information (Visual Subset). 
    In contrast, the evaluation task requires **Fine-Grained Relevance Reasoning**—determining the precise match level (L1--L4) between a query and an item. This requires specific domain knowledge, logical deduction, and strict adherence to annotation rules.
    The capability to \textit{identify} a problem as "hard" or "visually necessary" does not guarantee the capability to \textit{solve} it correctly according to complex guidelines. The cognitive demands of these two processes are largely orthogonal.

    \item \textbf{Ensemble Dilution:} As described in Eq.~2 and Eq.~3, our pipeline relies on \textbf{consensus mechanisms} (e.g., majority voting) involving diverse model architectures (e.g., Qwen, Llama, DeepSeek). A sample is retained only if it aligns with the consensus, effectively "diluting" the specific idiosyncrasies or hallucinations of any single model (including Qwen).
\end{enumerate}

In conclusion, while we cannot entirely rule out minor implicit biases, the structural design of RAIR—emphasizing task distinction and multi-model consensus—minimizes the impact of model-specific priors on the final leaderboard rankings.

\end{document}